\documentclass[journal,twocolumn]{IEEEtran}
\usepackage{amsmath}
\usepackage{amssymb}
\usepackage{graphicx}
\usepackage{tikz}
\usepackage{tkz-tab}
\usetikzlibrary{patterns}  
\usepackage{pgfplots}
\usepackage{booktabs}
\usepackage{floatflt}
\usepackage{enumerate}
\usepackage{psfrag}	
\usepackage{array}
\usepackage{multirow,hhline}
\usepackage{exscale}
\usepackage{color}
\usepackage{colortbl}
\usepackage{epsfig}
\usepackage{bm}
\usepackage{array}
\usepackage{subfigure}
\usepackage{nicefrac,xfrac}
\usepackage{siunitx}
\usepackage[ruled,vlined]{algorithm2e}

\usepackage{lettrine}
\usepackage{marginnote}

\usepackage[noadjust]{cite}

\newtheorem{theorem}{Theorem}

\newcommand{\RN}[1]{%
	\textup{\uppercase\expandafter{\romannumeral#1}}%
}

\DeclareMathOperator*{\argminA}{arg\,min} 

\begin{document}
	\bstctlcite{IEEEexample:BSTcontrol}
	\IEEEoverridecommandlockouts
	\title{Deep Unfolding of Iteratively Reweighted ADMM for
		Wireless RF Sensing}
	\author{{\small
			\IEEEauthorblockN{Udaya S.K.P. Miriya Thanthrige, Peter Jung, and Aydin Sezgin}}\\
		\thanks{{\small The work of U. S. K. P. M. Thanthrige and A. Sezgin is funded by the Deutsche Forschungsge-meinschaft (DFG, German Research Foundation)  Project--ID287022738 TRR 196 (S02) and the work of P. Jung is funded by the German Federal Ministry of Education and Research (BMBF) in the framework of the international future AI lab {``}AI4EO--Artificial Intelligence for Earth Observation: Reasoning, Uncertainties, Ethics and Beyond{''} (Grant number: 01DD20001).}} \thanks{{\small U. S. K. P. M. Thanthrige and A. Sezgin are with the Institute of Digital Communication Systems (DCS), Ruhr University Bochum (RUB), 44801 Bochum, Germany. Email: \{udaya.miriyathanthrige, aydin.sezgin\}@rub.de.}} \thanks{{\small P. Jung is with the Institute of Communications and Information Theory, Technical University Berlin, 10587 Berlin, Germany and Data Science in Earth Observation, Technical University of Munich, 82024 Taufkirchen/Ottobrunn, Germany. Email: peter.jung@tu-berlin.de.}}
	}
	
	\IEEEpubid{\begin{minipage}{\textwidth}\ \centering \textbf{“This work has been submitted for possible publication. Copyright may be transferred without notice, after which this version may no longer be accessible.”.}\end{minipage}
	}

	\maketitle

	\begin{abstract}
		We address the detection of material defects, which are inside a layered material structure using compressive sensing based multiple-input and multiple-output (MIMO) wireless radar. Here, the strong clutter due to the reflection of the layered structure’s surface often makes the detection of the defects challenging. Thus, sophisticated signal separation methods are required for improved defect detection. In many scenarios, the number of defects that we are interested in is limited and the signaling response of the layered structure can be modeled as a low-rank structure. Therefore, we propose joint rank and sparsity minimization for defect detection. In particular, we propose a non-convex approach based on the iteratively reweighted nuclear and $\ell_1-$norm (a double-reweighted approach) to obtain a higher accuracy compared to the conventional nuclear norm and $\ell_1-$norm minimization. To this end, an iterative algorithm is designed to estimate the low-rank and sparse contributions. Further, we propose deep learning to learn the parameters of the algorithm (i.e., algorithm unfolding) to improve the accuracy and the speed of convergence of the algorithm. Our numerical results show that the proposed approach outperforms the conventional approaches in terms of mean squared errors of the recovered low-rank and sparse components and the speed of convergence.
	\end{abstract}
	
	\begin{IEEEkeywords}
		Algorithm~unfolding, Clutter~suppression, Defects~detection, Compressive~sensing, Reweighted~norm
	\end{IEEEkeywords}

	%
	\IEEEpeerreviewmaketitle

	\section{Introduction}
	\label{sec:intro}
	\IEEEPARstart{N}{on}-destructive testing (NDT) is important in many areas such as remote sensing, security, and many more \cite{gholizadeh2016review, hillger2020toward, CHOPARD2021102473}. Further, NDT plays an important role in many industrial applications because it does not cause damage to the material under test. The electromagnetic (EM) waves-based remote sensing has many potential applications such as behind the wall object identification \cite{8234683}, multi-layer target detection \cite{8903033}, material characterization \cite{9178955}, defect detection \cite{CHOPARD2021102473,ZAHRAN201326, stoik2008nondestructive,UNNIKRISHNAKURUP2020102367} and many more. In EM waves-based detection of objects\slash defects which are behind or inside a layered structure, the EM waves that reflect from the object\slash defect are analyzed. Here, one major challenge is the presence of strong unwanted reflections, i.e., clutter \cite{8234683,huang2009uwb}. In the EM waves-based defect detection, the main source of the clutter is the reflection from the surface of the layered material structure.
	
	The state-of-the-art clutter suppression methods such as background subtraction (BS) and subspace projection (SP) \cite{khan2010background} are not able to suppress the clutter in the context of object\slash defect detection. 
	This is due to the fact in the BS requires the reference data of the scene and this reference data is not available most of the time. Moreover, in the SP prior
	knowledge is required to determine the perfect threshold for clutter removal. Further, clutter suppression becomes even more challenging if objects and clutter are closely located. This occurs regularly in the detection of 
	defects which are inside a layered structure. Then, due to the small delay spread between the signaling responses of defects and clutter superimpose each other. In order to overcome these challenges, advanced signal processing methods are required for clutter suppression. 	\IEEEpubidadjcol  \IEEEpubidadjcol
	
	Note that, in many scenarios, the number of defects is limited. Therefore, the signaling response of the defect is sparse in nature. By exploring this, compressive sensing (CS) \cite{1614066} based approaches have shown promising results in object\slash defect detection with clutter \cite{ 8234683,huang2009uwb}. Also, the CS-based approaches do not require full measurement data set, which results in fast data acquisition and less sensitivity to sensor failure, wireless interference and jamming. In CS based approaches, it is considered that the clutter resides in a low-rank subspace and the response of the objects is sparse \cite{8234683, huang2009uwb}. To this end, for the general data acquisition model the received data vector~$\bm{y}~\in \mathbb{C}^{K}$ is modelled as a combination of low-rank matrix~$\bm{L}~\in \mathbb{C}^{M\times N}$ and a sparse matrix~$\bm{S}~\in \mathbb{C}^{M\times N}$ with $M \leq N$
	\begin{equation}
		\bm{y}=\bm{A}_l\text{vec}(\bm{L})+\bm{A}_s\text{vec}(\bm{S}) + \bm{n}.
		\label{eq:SM_G}
	\end{equation}
	Here,
	$\bm{A}_l,~\bm{A}_s,~\in~\mathbb{C}^{K\times MN}$~with~$K~\ll MN$,~$\bm{n}~\in\mathbb{C}^{K}$~are compression operators and measurement noise, respectively. Further, $\text{vec}(\cdot)$ denotes the vectorization operator, which converts a matrix to a vector by stacking the columns of the matrix. Given the received data vector $\bm{y}$ our aim to estimate the signals of interest  $(\bm{L}~\text{and}~\bm{S})$ using a few number of linear measurement. In this problem, the compression ratio is defined as $K/MN$. Note that the above problem is also known as robust principal component analysis (RPCA) \cite{candes2011robust}. The problem given in~\eqref{eq:SM_G} has different types (a) Standard/classical RPCA in which both $\bm{A}_l$ and $\bm{A}_s$ in~\eqref{eq:SM_G} are identity matrices \cite{candes2011robust}, (b) The matrices $\bm{A}_l=\bm{A}_s=\bm{A}$ and $\bm{A}$ is a selection operator which selects a random
	subset of size $K$ from $MN$ entries~\cite{tang2011constrained}, (c) Both $\bm{A}_l$ and $\bm{A}_s$ are linear operators which map the vector space $MN$ to the vector space $K$~\cite{tang2011constrained}. In this work, we focus on (c) and we consider two cases where $\bm{A}_l=\bm{A}_s=\bm{A}$ and $\bm{A}_l\neq\bm{A}_s$. Estimating a matrix with the lowest rank and the sparsest matrix from the compressed observations are NP-hard problems and thus difficult to solve. To this end, convex relaxations of sparsity ($\ell_0-$norm: the number of nonzero components) and rank in terms of $\ell_1-$norm of a matrix (absolute sum of
	elements) and nuclear norm of a matrix (sum of singular values of a matrix) are utilized, respectively \cite{ bruckstein2009sparse, cai2010singular,fazel2001rank}.
	
	Next, we briefly discuss the recovery guarantees of the RPCA problem. Here, we focus on the standard RPCA problem i.e., both $\bm{A}_l$ and $\bm{A}_s$ in~\eqref{eq:SM_G} are identity matrices. We consider above scenario because it is well studied in the literature and well understood. So we briefly discuss when the separation of low-rank matrix $\bm{L}$ and sparse matrix $\bm{S}$ is possible. Based on~\cite{candes2011robust}, if $\bm{L}$ is sufficiently low-rank but not sparse and $\bm{S}$ is sufficiently sparse but not low-rank, the matrices $\bm{L}$ and $\bm{S}$ can be estimated exactly with a high probability of success. Here, to solve the RPCA problem 
		convex relaxations of sparsity and rank in terms of $\ell_1-$norm of a matrix and nuclear norm of a matrix are utilized \cite{candes2011robust}. Let $\bar{K}=$ \text{max}($N,M$), $K=$ \normalfont{min}($N,M$) and positive constants $c_o$, $p_s$ and  $p_r$. We consider the following theorem from \cite{candes2011robust}.
		\begin{theorem}
			It is possible to recover $\bm{L}$ and $\bm{S}$ with a probability at least $1-c\bar{K}^{-10}$ when $\text{rank} (\bm{L}$)$ \ \le \ p_r K (\mu)^{-1} (log(\bar{K}))^{-1}$ $\text{and}$
			$\parallel \bm{S} \parallel_0 \ \le \ p_s \bar{K} K$.
			\label{theorem1}
	\end{theorem}
	
	Here, $\mu$ is the incoherence condition parameter of the low-rank matrix $\bm{L}$ as defined in \cite{candes2011robust}. As discussed in \cite{candes2009exact,gross2011recovering,candes2010power} when the incoherence condition parameter $\mu$ is small, the singular value vector of the matrix $\bm{L}$ is spread out. The main message of Theorem \ref{theorem1} is that to recover the $\bm{L}$ and $\bm{S}$ successfully, the rank of the matrix $\bm{L}$ should not be too large and the matrix $\bm{S}$ should be reasonably sparse. That means that the rank of the matrix $\bm{L}$ and sparsity of the matrix $\bm{S}$ is bounded.
	
	To this end, we have shortly discussed when the separation of low-rank and sparse matrices make sense. Next, we shortly discuss what is the lowest possible mean squared error of the low-rank and sparse matrix recovery that can be achieved. For that we have chosen the Cram\'{e}r–Rao bound (CRB) analysis of the RPCA given in~\cite{tang2011constrained}. For RPCA problem with $\bm{A}_l=\bm{A}_s=\bm{A}$ in~\eqref{eq:SM_G}, the CRB of unbiased estimation for $\bm{L}$ and $\bm{S}$ is derived in~\cite{tang2011constrained}. Here, similar to \cite{candes2011robust}, the RPCA problem is solved using the $\ell_1-$norm and nuclear norm minimization.
		To this end, let the estimated low-rank and sparse matrices be given by $\hat{\bm{L}}$ and $\hat{\bm{S}}$, respectively. Now, similar to Theorem \ref{theorem1}, it is assumed that the rank of the matrix $\bm{L}$ and the sparsity of the matrix $\bm{S}$ is bounded as given below.
		\begin{equation}
			\begin{aligned}
				\text{rank} (\bm{L}) =r \ &\le l,\\
				\parallel \bm{S} \parallel_0 \ &\le s.
			\end{aligned}
		\end{equation} 
		Here, $l$ and $s$ are positive constants. Let the received data vector $\bm{y}$ in~\eqref{eq:SM_G} follows $\bm{y} \sim \mathcal{N}(\bm{A}\text{vec}(\bm{L}+\bm{S}),\,\sigma^{2}\bm{I}_K)$. Now, CRB  of unbiased estimation for $\bm{L}$ and $\bm{S}$ is bounded by\footnote{Here, $\bm{A}$ is assumed to be a selection operator which selects a random subset of size $K$ from $MN$ entries. Since this is closest matching CRB to our model given in~\eqref{eq:SM_G}, we have considered this formulation as a benchmark.}
	\begin{equation}
		\begin{aligned}
			\left\{s-N_0+\dfrac{1}{3}\dfrac{KN_0}{K-s}+\dfrac{2}{3}\dfrac{MNN_0}{K-s}\right\}\sigma^{2} \le \text{CRB}(\bm{L},\bm{S})\le \\ \left\{s-N_0+\dfrac{3KN_0}{K-s}+\dfrac{2MNN_0}{K-s}\right\}\sigma^{2}.
			\label{eq:crb}
		\end{aligned}
	\end{equation}
	Here, $\bm{I}_K$ is the identity matrix with a dimension of $K \times K$ and  $N_0=(M+N)r-r^2$.
	
	Next, we are going to discuss the disadvantages of convex relaxation of RPCA and the alternative approaches instead of the convex relaxation. In general, the $\ell_1$-norm minimization and the $\ell_0$-norm minimization have different solutions. Further, the $\ell_1$-norm minimization is a loose approximation of the $\ell_0$-norm minimization. Due to this reason, alternative approaches such as the weighted $\ell_1-$norm minimization have been considered for improved performance over the $\ell_1$-norm minimization \cite{candes2008enhancing }.
	
	Note that in many applications, the important properties are preserved by the large coefficients\slash singular values of the signal. However, the $\ell_1-$norm\slash nuclear norm minimization algorithms shrink all the coefficients\slash singular values with the same threshold. Thus, to avoid this weakness, we should shrink less the larger coefficients\slash singular values. Therefore, in both $\ell_1-$norm and nuclear norm minimization, the weighted $\ell_1-$norm and weighted nuclear norm minimization have been considered \cite{candes2008enhancing, mohan2010reweighted, gu2017weighted}. Note that, the non-convex approaches, i.e., reweighted nuclear norm and $\ell_1$-norm minimization have shown better performance over the convex relaxations by providing tighter characterizations of rank and sparsity.
		For instance, studies \cite{candes2008enhancing,daubechies2010iteratively} have shown that the reweighted $\ell_1$-norm minimization outperforms the $\ell_1$-norm minimization. Similarly, studies \cite{mohan2010reweighted, gu2017weighted} have shown that the reweighted nuclear norm outperforms the nuclear norm minimization. Although reweighted approached like the reweighted $\ell_1$-norm minimization outperforms the convex $\ell_1$-norm minimization, still the reweighted $\ell_1$-norm minimization and their convergence has not been fully studied \cite{Zhao2012}.
		
				 To this end, we formulate an optimization problem based on the reweighted nuclear norm and reweighted $\ell_1-$norm minimization to jointly estimate the low-rank matrix and the sparse vector from few compressive measurements. Note that our iterative algorithm is based on the alternating direction method of multipliers (ADMM) \cite{7889039}. Then, the estimated sparse vector is used to identify defects. To the best of our knowledge, the full doubly-reweighted approach, i.e., wrt. nuclear norm and $\ell_1-$norm has not yet been studied in the literature for the compressive case.
				 
	In iterative algorithms, the accuracy of the recovered signal component and the convergence rate
	depends on the proper selection of parameters (e.g.,~regularization\slash thresholding\slash denoising~parameters). Generally, parameters are chosen by handcrafting, and it is a time-consuming task. In this context, machine learning-based parameter learning using the training data has shown promising results in many applications such as sparse vector recovery \cite{gregor2010learning,9023989,9414910} and image processing \cite{8950351}. This approach is known as algorithm unrolling\slash unfolding. An interesting overview of the algorithm unfolding can be found in \cite{9363511}. The authors in \cite{gregor2010learning} unfolded the iterative shrinkage and thresholding algorithm (ISTA). It is called as learned ISTA (LISTA) and the LISTA converges twenty times faster than the ISTA.
	
	In this work, we propose deep learning-based parameter learning to improve the accuracies of the recovered low-rank and sparse components and the convergence rate of the ADMM based iterative algorithm. In the following, we emphasize the differences of our approach with the existing low-rank plus sparse recovery approaches. Most of the work in the literature focus on the standard RPCA where both $\bm{A}_l$ and $\bm{A}_s$ in~\eqref{eq:SM_G} to be identity matrices  \cite{gu2014weighted,gu2017weighted}. Further, most of the time RPCA problems are solved by using the convex relaxation, i.e., nuclear norm and $\ell_1-$norm minimization or with the single reweighting, i.e., either reweighted $\ell_1-$norm or reweighted nuclear norm \cite{gu2014weighted,gu2017weighted,8836615,8683030}. In this work, we study the joint reweighted nuclear norm and reweighted $\ell_1-$norm (full doubly-reweighted) compressive sensing data acquisition model. In the context of the algorithm unfolding for the RPCA, the convolutional robust principal component analysis~(CORONA)~\cite{8836615} and \cite{8683030} are the closest work to our work. There are fundamental differences between our work and \cite{8683030,8836615}. (a) Our methodology is fundamentally different from the methodology that both \cite{8683030,8836615} have used to solve the RPCA. In more details, both \cite{8683030,8836615} considered the standard convex relaxation ($\ell_{1,2}-$norm and nuclear norm) to solve the RPCA problem, while we propose the reweighted $\ell_1-$norm and reweighted nuclear norm.
	
	 Further, our iterative algorithm to solve RPCA is based on ADMM while iterative algorithm in \cite{8683030,8836615} based on fast ISTA (FISTA). The motivation to propose ADMM over ISTA/FISTA for RPCA is as follows. As shown in \cite{candes2011robust,yuan2009sparse} for RPCA, the ADMM based approach is able to achieve the desired solution with a good recovery error with a small number of iteration for a wide range of applications. Also, it is shown that ADMM converges faster than ISTA/FISTA. As an example for $\ell_1$-norm minimization, in \cite{tao2015convergence} it is shown that the ADMM achieves the desired solution with fewer iterations than ISTA/FISTA. (b) Different from \cite{8683030,8836615} our focus is on defect detection based on the stepped-frequency continuous (SFCW) radar while \cite{8683030,8836615} focus on ultrasound imaging application. 
	Further, measurement data of \cite{8683030,8836615} have considered that both $\bm{A}_l$ and $\bm{A}_s$ in~\eqref{eq:SM_G} are identity matrices. However, in this work, we consider the joint reweighted nuclear norm and reweighted $\ell_1-$norm (full doubly-reweighted) with a compressive sensing data acquisition model i.e., both $\bm{A}_l$ and $\bm{A}_s$ in~\eqref{eq:SM_G} are not identity matrices. Further, we have studied the performance of our approach with a generic real-value Gaussian model and a complex-valued SFCW radar model for different compression ratios.
	
	In addition to that, CORONA is based on a convolutional neural network while our approach is based on a dense neural network. The main motivation to prefer a dense neural network over a convolutional neural network is that the system model of the defect detection based on the SFCW radar does not include the convolutional operation. Therefore, we consider a dense DNN which match to our system model more than the convolutional DNN. In addition to that in the compressive sensing data acquisition model, the received data vector $\bm{y}$ in~\eqref{eq:SM_G} has a much lower dimension than the low-rank and sparse matrices $\bm{L}$ and $\bm{S}$ i.e., $K~\ll MN$. In the estimation of $\bm{L}$ and $\bm{S}$ from $\bm{y}$, lower dimension to higher dimension mapping is required. Here, we experience that the performance of the convolution DNN is limited. This is due to the fact that, as the compression ratio increases more zero padding is required. Another advantage of the dense DNN is the simplicity of initialization of the weights of the DNN. Here, it is possible to initialize the weights of the DNN by using the measurement matrices $\bm{A}_l$ and $\bm{A}_s$. Also, for some applications, there is no such requirement to train the weights of the DNN that are initialized based on the measurement matrices. Here, only training the parameters of the iterative algorithm is sufficient to get an improved performance. However, with a convolution DNN this may not be straightforward as a dense DNN. Because finding a relationship between the measurement matrices and the weights of the convolutional DNN is not straightforward. 
		
		Also, in CORONA \cite{8836615}, a custom complex-valued convolution layers and singular value decomposition operations are utilized. Note that in our work, we have implemented a dense DNN which supports complex-valued data in linear layers and singular value decomposition (SVD) operation.  
	\subsection{Contribution}
	\label{sec:Contribution}
	The contributions of this work are summarized as follows:
	\begin{itemize}
		
		\item We propose the non-convex fully double-reweighted approach i.e., both reweighted $\ell_1-$norm and reweighted nuclear norm simultaneously to solve the RPCA problem in the compressive sensing data acquisition model which reflects more the practical problem at hand.
		\item We propose an iterative algorithm based on ADMM to estimate the low-rank and sparse components jointly. 
		
		\item We propose a deep neural network (DNN) to learn the parameters of the iterative algorithm (i.e., algorithm unfolding) from the training data.
		
		\item We consider two types of data to evaluate our approach. As a practical application, the defect detection by SFCW radar from compressive measurements with $\bm{A}_l\neq\bm{A}_s$ is used. In addition to that, for a standard benchmark, a generic Gaussian data acquisition model with $\bm{A}_l=\bm{A}_s=\bm{A}$ is used. Further, we compare the performance of the proposed DNN based approach with the standard convex approach (i.e., nuclear norm and $\ell_1-$norm minimization) and the untrained ADMM based iterative algorithm for different compression ratios. 
		In both cases, our numerical results show that the proposed approach outperforms the conventional approaches in terms of mean squared errors of the recovered low-rank and sparse components and the speed of convergence.
	\end{itemize}

	The remainder of the paper is organized
	as follows. We introduce the SFCW radar-based defect detection and the low-rank plus sparse recovery with reweighting in Section~\ref{sec:System_Model}. In Section~\ref{sec:Unfolding_ADMM} we discuss the DNN-based low-rank plus sparse recovery algorithm unfolding. In Section~\ref{sec:results}, we provide an evaluation of the proposed DNN-based low-rank plus sparse recovery algorithm unfolding approaches and provide interesting insights. Section~\ref{sec:concl} concludes the~paper. 
	\section{System Model}
	\label{sec:System_Model}
	First, we briefly present the system model of the mono-static SFCW radar-based defect detection. Next, we are going to discuss the ADMM  based iterative algorithm for the low-rank plus sparse recovery.

	\subsection{SFCW radar based defect detection}
	\label{sec:SFCWSystem_Model}
\begin{figure}[!t]
		\centering
		\includegraphics[width=0.90\linewidth]{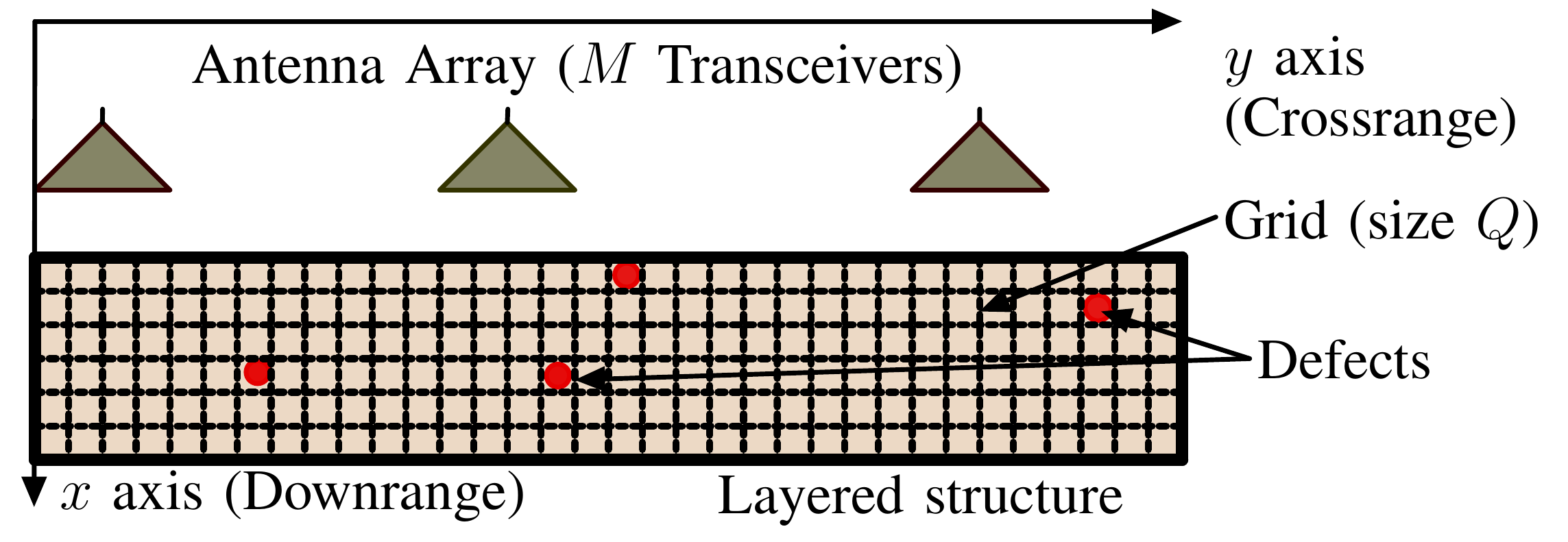}
		\caption{Getting the measurements of a single-layered material structure using a SFCW radar with $M$ transceivers.}
		\label{fig:fig1}
	\end{figure}

	We consider a SFCW radar with $M$ transceivers which are placed in parallel to the single-layered material structure while maintaining an equal distance between transceivers as shown in Fig.~\ref{fig:fig1}. In SFCW radar, each transceiver transmits a stepped-frequency signal containing $N$ frequencies which are equally spaced over the bandwidth of $B$ Hz. To this end, the received signal corresponding to all $M$ transceivers and $N$ frequencies $\bm{Y} \in\mathbb{C}^{M \times N}$  are given by 
	\begin{equation}
		\bm{Y}= \bm{Y}^l+ \bm{Y}^d+ \bm{Z}.
	\end{equation} 
 Note that, $\bm{Y}$ consists of two main components, the reflection of the layered material structure ($\bm{Y}^l$) and the reflection of the defects ($\bm{Y}^d$). Here, $\bm{Z}$ is the additive Gaussian noise matrix. Next, we discuss in details the modeling of the received signal of the defects by using the propagation time delay. To this end, the scene shown in Fig.~\ref{fig:fig1} is hypothetically partitioned into a rectangular grid of size~$Q$. Suppose that the round-travel time of the signal from the \mbox{$m$-th}~antenna location to the \mbox{$p$-th}~defect and back is given by $\tau_{m, p}$. Then, the received signal of the defects in \mbox{$m$-th}~transceiver corresponding to \mbox{$n$-th}~frequency band $(f_{n})$ $y_{m,n}^d$ is given by
	\begin{equation}
		y_{m,n}^d=\sum\limits_{p=1}^{P}\alpha_{p}~\text{exp}(-j2\pi f_{n}\tau_{m, p}).
		\label{eq:SS_eq_rx_obj_vec}
	\end{equation}
	Here, $j\!=\!\sqrt{-1}$, $\alpha_{p}~\in\mathbb{C}$~is the complex signal strength of the~\mbox{$p$-th}~defect and~$P$~is the total number of defects. To this end, $\text{vec}(\bm{Y}^d)~\in\mathbb{C}^{MN \times 1}$ is given by $\text{vec}(\bm{Y}^d)=\bm{D}\bm{s}$, where vector~$\bm{s}~\in\mathbb{C}^{Q \times 1}$ contains all the $\alpha_{p}$ values of the defects. Since there are $P$ defects, the vector~$\bm{s}$ only contains $P$ \mbox{non-zero} entries. The matrix~$\bm{D}$ is given by $[(\bm{D}_{1})^T,...,(\bm{D}_{m})^T,...,(\bm{D}_{M})^T]^T \in\mathbb{C}^{MN \times Q}$. Note that, the \mbox{$(n,q)$-th}~element of the matrix~$\bm{D}_{m} \in \mathbb{C}^{N \times Q}$ is given by $\text{exp}(-j2\pi f_{n}\tau_{m,q})$, where $\tau_{m,q}$ is the propagation time delay between the \mbox{$m$-th}~antenna to the \mbox{$q$-th}~grid location. Note that, we assume that the propagation time delays of the defects are exactly matched with the propagation time delays of the grid locations.
	\subsection{Compressed sensing (CS) approach}
	\label{sec:SFCWCS}
	In the compressed sensing (CS) setup only a subset of antennas\slash frequencies are available or selected. Now, the reduced data vector~$\bm{y}_{cs} \in\mathbb{C}^{K \times 1}$ of size $K~(\ll MN)$ is given by
	\begin{equation}
		\begin{aligned}
			\bm{y}_{cs}=\bm{\Phi}\left(\text{vec}(\bm{Y})\right)
			=\bm{\Phi}\text{vec}(\bm{Y}^l)+\bm{\Phi}\bm{D}\bm{s}+\bm{\Phi}\text{vec}(\bm{Z}),
		\end{aligned}
		\label{eq:SS_eq_rx_all_cs}
	\end{equation} 
	where $\bm{\Phi} \in\mathbb{R}^{K \times MN} $ is the selection matrix. The matrix~$\bm{\Phi}$ has a single \mbox{non-zero} element of value $1$ in each row to indicate the selected frequency of a particular antenna if that antenna is selected. Here, our main objective is to recover $\bm{Y}^l$~and~$\bm{s}$ from the reduced data vector~$\bm{y}_{cs}$ using the low-rank plus sparse recovery approach as detailed below.
	\subsection{Low-rank-plus-sparse recovery algorithm}
	From now on we consider the general data acquisition model given in~\eqref{eq:SM_G} in Section \ref{sec:intro} i.e., $\bm{y}=\bm{A}_l\text{vec}(\bm{L})+\bm{A}_s\text{vec}(\bm{S}) +\bm{n}$. Note that, the SFCW radar model given in~\eqref{eq:SS_eq_rx_all_cs} is mapped to the generic measurement model by considering $\bm{A}_s\!=\!\bm{\Phi}\bm{D}$, $\bm{A}_l\!=\!\bm{\Phi}$,~$\bm{Y}^l\!=\!\bm{L}$,~$\bm{s}=\text{vec}(\bm{S})$~and~$\bm{y}_{cs}~=~\bm{y}$, respectively. Our objective is to recover the low-rank matrix~$\bm{L}$ and the sparse matrix~$\bm{S}$ from the compressive measurements~$\bm{y}$. Thus, the estimation of $\bm{L}$ and $\bm{S}$ from $\bm{y}$ is done by minimizing rank and the sparsity ($\ell_0-$norm). Note that, rank and $\ell_0-$norm
	minimization problems are usually \mbox{NP-hard}. Thus,
	one may use convex relaxations based on the nuclear norm of a matrix and $\ell_1-$norm of a matrix as follows
	\begin{equation}
		\label{eq:lowrank_sparse_A_cov}
		\begin{aligned}
			\left \{\hat{\bm{L}}, \hat{\bm{S}} \right\} &=\argminA_{\bm{L},  \ \bm{S}}  \lambda_l\left\Vert\bm{L}\right\Vert_* + \lambda_s\left\Vert \bm{S} \right\Vert_1, \\
			& \quad \  \ \text{s.t.} \ \left\Vert \bm{y}-\bm{A}_s\text{vec}(\bm{S})-\bm{A}_l\text{vec}(\bm{L}) \right\Vert^2_2 \  \le \ \epsilon. 
		\end{aligned}
	\end{equation}
	Here, $\lambda_l$, $\lambda_s$ are regularization parameters and  $\epsilon$ is a small positive constant (noise bound). Further, $\left\Vert \cdot \right\Vert_1 $, $\left\Vert\cdot\right\Vert_*$ are the $\ell_1-$norm and nuclear norm of a matrix, respectively. The resulting convex problems, i.e., $\ell_1-$norm and nuclear norm minimization are 
	well studied in the literature and there are several non-convex approaches to improve over the standard convex relaxation. One well-known approach is iterative reweighting of the $\ell_1-$norm \cite{candes2008enhancing, 5419071,9023989} and nuclear norm \cite{gu2017weighted,6858068, Fazel2003,7362209}. 
	Alternating direction method of multipliers (ADMM) is used to solve the problem given in~\eqref{eq:lowrank_sparse_A_cov} \cite{7889039}. First, we formulate the problem given in~\eqref{eq:lowrank_sparse_A_cov} based on the ADMM approach, and then we introduce the non-convex double-reweighted approach i.e., both reweighted $\ell_1-$norm and reweighted nuclear norm simultaneously. Let the signal component value of $\bm{S}$ and $\bm{L}$ at the \mbox{$t-$th}~iteration be denoted as $\cdot^t$. Now, based on the ADMM $\bm{S}$ and $\bm{L}$ are estimated by  
	\begin{equation}
		\label{eq:lowrank_sparse_A_ADMM_L}
		\begin{aligned}
			\bm{L}^{t+1} &= \argminA_{\bm{L}}  \  \lambda_l\left\Vert \bm{L} \right\Vert_{\star} + \\ &\ \dfrac{\rho}{2}  \left\Vert \bm{A}_s\text{vec}\left(\bm{S}^t\right)  +\bm{A}_l\text{vec}(\bm{L})-\bm{y} + \dfrac{1}{\rho} \bm{u}^t  \right\Vert_2^2,
		\end{aligned}
	\end{equation} 
	\begin{equation}
		\label{eq:lowrank_sparse_A_ADMM_S}
		\begin{aligned}
			\bm{S}^{t+1} &= \argminA_{\bm{S}} \  \lambda_s \left\Vert \bm{S} \right\Vert_1  + \\ &\ \dfrac{\rho}{2} \left\Vert \bm{A}_s\text{vec}\left(\bm{S}\right)  +\bm{A}_l\text{vec}\left(\bm{L}^{t+1}\right)-\bm{y} + \dfrac{1}{\rho} \bm{u}^t  \right\Vert_2^2,
		\end{aligned}
	\end{equation}
	\begin{equation}
		\label{eq:lowrank_sparse_A_ADMM_U}
		\bm{u}^{t+1} =\bm{u}^{t}+\rho \left(\bm{A}_s\text{vec}\left(\bm{S}^{t+1}\right) +\bm{A}_l\text{vec}\left(\bm{L}^{t+1}\right)-\bm{y} \right).
	\end{equation}
	Here, $\bm{u}$, $\rho >0$ are auxiliary variables and a penalty factor. Let $\bm{\sigma}(\bm{L})\!=\![\sigma_1,...\sigma_m,...,\sigma_M]\in~\mathbb{R}^{M}$ be the singular values~of the matrix~$\bm{L}$. Now,  nuclear norm of the matrix $\bm{L}$ is given by $\left\Vert \bm{L} \right\Vert_{\star}=\left\Vert \bm{\sigma}(\bm{L}) \right\Vert_{1}$. Now, we are going to introduce the weighted $\ell_1-$norm  and weighted nuclear norm to the sub-problems given in~\eqref{eq:lowrank_sparse_A_ADMM_L} and \eqref{eq:lowrank_sparse_A_ADMM_S} as follows.	
	\begin{equation}
		\label{eq:lowrank_sparse_A_ADMM_LRW}
		\begin{aligned}
			\bm{L}^{t+1} 
			&= \argminA_{\bm{L}}  \  \lambda_l\left\Vert \bm{w}_l^t\odot\bm{\sigma}(\bm{L}) \right\Vert_1+ \\ &\   \dfrac{\rho}{2}  \left\Vert \bm{A}_s\text{vec}\left(\bm{S}^t\right) +\bm{A}_l\text{vec}(\bm{L})-\bm{y} + \dfrac{1}{\rho} \bm{u}^t  \right\Vert_2^2,
		\end{aligned}
	\end{equation} 
	\begin{equation}
		\label{eq:lowrank_sparse_A_ADMM_SRW}
		\begin{aligned}
			\bm{S}^{t+1} &= \argminA_{\bm{S}} \  \lambda_s \left\Vert \bm{w}_s^t\odot\bm{S} \right\Vert_1 + \\ &\  \dfrac{\rho}{2} \left\Vert \bm{A}_s\text{vec}(\bm{S})  +\bm{A}_l\text{vec}\left(\bm{L}^{t+1}\right)-\bm{y} + \dfrac{1}{\rho} \bm{u}^t  \right\Vert_2^2.
		\end{aligned}
	\end{equation}
	The operator $\odot$ denotes element-wise multiplication. Here, $\bm{w}_l^t~\in~\mathbb{R}^{M}$ and  $\bm{w}_s^t~\in~\mathbb{R}^{MN}$~are \mbox{non-negative} weight vectors. To this end, $\bm{w}_l^t$ and $\bm{w}_s^t$ are calculated based on the previous estimation of the $\bm{L}$ and $\bm{S}$ i.e., $\bm{L}^{t}$ and $\bm{S}^{t}$.
	\begin{equation}
		\label{eq:WL}
		\bm{w}_l^t =g_l\left(\bm{\sigma}\left(\bm{L}^{t} \right)\right)~\text{and}~\bm{w}_s^t =g_s\left(|\bm{S}^{t}|\right).
	\end{equation}
	\noindent Here, $g_l(\cdot)$ and $g_s(\cdot)$ are the decay functions which are used to calculate the weights. There are several decay functions are proposed in the literature and an overview
	of many known decay functions for the nuclear
	norm is given in \cite{7362209}. In this work, we consider the \mbox{element-wise} \mbox{soft-thresholding} \cite{9023989} and \mbox{element-wise} singular value \mbox{soft-thresholding} (i.e., \mbox{element-wise} \mbox{soft-thresholding} on the singular value of a matrix) \cite{peng2014reweighted} as proximal operators of weighted $\ell_1-$norm and weighted nuclear norm, respectively. Now, $\bm{L}^{t+1}$ and $\bm{S}^{t+1}$ are given by
	\begin{equation}
		\label{eq:lowrank_sparse_A_WADMM_SVD}
		\bm{L}^{t+1} 
		= \text{SVT}_{\bm{\lambda}_{LT}}\left(\bm{A}_l^	\ast\left(\bm{y} -\bm{A}_s\text{vec}\left(\bm{S}^{t}\right)+\dfrac{\bm{u}^t}{\rho}   \right)\right),	
	\end{equation}
	\begin{equation}
		\label{eq:lowrank_sparse_A_WADMM_ST}
		\bm{S}^{t+1} 
		= \text{ST}_{\bm{\lambda}_{ST}}\left( \bm{A}_s^	\ast\left(\bm{y} -\bm{A}_l\text{vec}\left(\bm{L}^{t+1}\right)+\dfrac{\bm{u}^t}{\rho} \right)\right),
	\end{equation}
	where  $\text{SVT}(\cdot)$ and $\text{ST}(\cdot)$ are the \mbox{element-wise} singular value \mbox{soft-thresholding} and \mbox{element-wise} \mbox{soft-thresholding} operators \cite{peng2014reweighted, 9023989}, respectively. Note that, $(\cdot)^\ast$ is a linear operator which back project the vector in to the target matrix subspace. There are two options for $(\cdot)^\ast$. (a) Hermitian transpose $(\cdot)^H$ as done in \cite{9023989}. (b) Moore–Penrose pseudo inverse $(\cdot)^\dagger$ as done in \cite{8234683}. Next, we are going to discuss the \mbox{element-wise} \mbox{soft-thresholding} in more details.
	\subsection{Element-wise soft-thresholding and singular value soft-thresholding}
	\label{EWST}
	Consider~\eqref{eq:lowrank_sparse_A_WADMM_ST} and let the intermediate value in the \mbox{$t+1-$th}~iteration as $\bm{S}_{I}^{t+1} \in \mathbb{C}^{M\times N}$ before the \mbox{soft-thresholding} be given by
	\begin{equation}
		\begin{aligned}
			\bm{S}_{I}^{t+1} = \bm{A}_s^\ast\left(\bm{y} -\bm{A}_l\text{vec}\left(\bm{L}^{t+1}\right)+\dfrac{1}{\rho}  (\bm{u})^t\right).
		\end{aligned}
	\end{equation}
	To estimate the sparse component in the~\mbox{$t+1-$th}~iteration $\bm{S}^{t+1}$, \mbox{element-wise} \mbox{soft-thresholding} is applied to each element of the matrix~$\bm{S}_{I}^{t+1}$ individually. To this end, the \mbox{element-wise} \mbox{soft-thresholding} operation of the \mbox{$m-$th}~row and \mbox{$n-$th}~column element of the matrix~$\bm{S}_{I}^{t+1}$ is defined as \footnote{Note that, here we consider the complex-version of the soft-thresholding.}
	\begin{equation}
		\text{ST}_{\lambda_{ST}^{m,n}}(s_{m,n}^{t+1})=\text{exp}(j\theta)~\text{max}\left(|s_{m,n}^{t+1}|-\lambda_{ST}^{m,n},0\right).
	\end{equation}
	Here, $\theta$ is the phase angle of the $s_{m,n}^{t+1}$ in radians. Here, $\lambda_{ST}^{m,n}$ is the threshold value related to the element $s_{m,n}^{t+1}$.
	Note that the threshold values ($\lambda_{ST}^{m,n}~\forall~ m,~n$) change from iteration to iteration. However, for better readability, we not include the index $t$. Here, a decay function is used to calculate the threshold value. In this work, we consider the common \mbox{log-determinant} heuristic \cite{candes2008enhancing} and exponential decay \cite{6858068,7362209} as decay
	functions. To this end, the \mbox{log-determinant} heuristic decay function is given by
	\begin{equation}  
		g_s(x)=g_{\text{log}}(x)=  \dfrac{1}{x+\gamma}.
	\end{equation}
	Similarly, the exponential heuristic decay function is given by
	\begin{equation}  
		g_s(x)=g_{\text{exp}}(x)=  \dfrac{1} {\gamma}\text{exp}\left(\dfrac{-x} {\gamma} \right).
	\end{equation}
	Here, $\gamma$ is a positive constant. To this end, the estimation of matrix~$\bm{S}$ in \mbox{$t+1-$th}~iteration, i.e., $\bm{S}^{t+1}$ is given by
	\begin{equation}
		\label{eq:lowrank_sparse_6b}
		\begin{aligned}
			\bm{S}^{t+1} &= \text{ST}_{\bm{\lambda}_{ST}^t}\left( \bm{S}_{I}^{t+1}\right),
		\end{aligned}
	\end{equation}
	where   $\bm{\lambda}_{ST}^t=[\lambda_{ST}^{1,1},...,\lambda_{ST}^{m,n},...,\lambda_{ST}^{M,N}]$ contains different threshold values for each element of $\bm{S}$ for the~\mbox{$t+1-$th}~iteration. These threshold values are derived based on the previous estimate of $\bm{S}$, i.e., $\bm{S}^{t}$. Thus, $\lambda_{ST}^{m,n}$ is given by
	\begin{equation}
		\label{eq:lowrank_sparse_6c}
		\begin{aligned}
			\lambda_{ST}^{m,n}=\lambda_{S}~g_s(|s_{m,n}^{t}|).
		\end{aligned}
	\end{equation}
	Here $\lambda_{S}$ is a positive constant and $s_{m,n}^{t}$ is the \mbox{$m-$th}~row and \mbox{$n-$th}~column element of the \mbox{$t-$th}~estimation of the sparse component $\left(\bm{S}^{t}\right)$. 
	The same concept is also applied to the singular value \mbox{soft-thresholding} which is used to solve the problem given in~\eqref{eq:lowrank_sparse_A_WADMM_SVD} as discussed next. In this work, we consider the same decay function for both sparsity and rank, i.e., $g_s(\cdot)=g_l(\cdot)$. First, consider~\eqref{eq:lowrank_sparse_A_WADMM_SVD} and let the intermediate value in~\mbox{$t+1-$th}~iteration, $\bm{L}_{I}^{t+1} \in \mathbb{C}^{M\times N}$ before applying the singular value \mbox{soft-thresholding} be given by
	\begin{equation}
		\begin{aligned}
			\bm{L}_{I}^{t+1} = \bm{A}_l^\ast\left(\bm{y} -\bm{A}_s\text{vec}\left(\bm{S}^{t}\right)+\dfrac{1}{\rho}  (\bm{u})^t\right).
		\end{aligned}
	\end{equation}
	\noindent The singular value decomposition (SVD) of the matrix~$\bm{L}_{I}^{t+1}\in \mathbb{C}^{M\times N}$ with $M \leq N$  is given as $\bm{L}_{I}^{t+1}=\mathbf{U}\boldsymbol \Lambda\mathbf{V}^{T}$. Here, $\mathbf{U}~\in \mathbb{C}^{M\times M}$ and $\mathbf{V}~\in \mathbb{C}^{N\times N}$ are the matrices of left and right singular vectors. $\mathbf{\Lambda}~\in~\mathbb{R}^{M \times N}$ is a rectangular diagonal matrix with $\bm{\sigma}(\bm{L}_{I}^{t+1})=[\sigma_1,...\sigma_m,...,\sigma_M]$ on the diagonal and zeros elsewhere. Next, the element-wise singular value \mbox{soft-thresholding} is applied to the singular values of the matrix~$\bm{L}_{I}^{t+1}$ to estimate the $\bm{L}^{t+1}$
	\begin{equation}
		\begin{aligned}
			\label{eq:lowrank_sparse_5bv2}
			\bm{L}^{t+1}&=\text{SVT}_{\bm{\lambda}_{LT}^t} \left( \bm{L}_{I}^{t+1}\right),\\ &= \mathbf{U}\text{diag} \left(\text{ST}_{\bm{\lambda}_{LT}^t}\left(\bm{\sigma}\left(\bm{L}_{I}^{t+1}\right)\right)\right)\mathbf{V}^{T}.
		\end{aligned}	
	\end{equation}
	Note that, here $\text{ST}_{\bm{\lambda}_{LT}^t}(\cdot)$ is the \mbox{element-wise} \mbox{soft-thresholding} operator and the operator $\text{diag}(\cdot)$ takes a vector as an input and returns a square diagonal matrix in which the main diagonal contains the vector elements and zeros elsewhere. Here, $\bm{\lambda}_{LT}^t=[{\lambda}_{LT}^1,...,{\lambda}_{LT}^m,...{\lambda}_{LT}^M]$ contains the different threshold values and ${\lambda}_{LT}^m$
	is calculated based on the singular values of the previous estimate of matrix~$\bm{L}$ as given below. 
	\begin{equation}
		\label{eq:lowrank_sparse_6b}
		\begin{aligned}
			{\lambda}_{LT}^m=\lambda_{L}~g_l(\sigma_m^t).
		\end{aligned}
	\end{equation}
	Here, $\sigma_m^t$ is the \mbox{$m-$th}~singular value of the $\bm{L}^{t}$ and $\lambda_{L}$ is a positive constant.
	\begin{algorithm}[h]
		{\fontsize{9.9pt}{14} \selectfont 
			\SetInd{0.2em}{0.4em}
			\SetKwInput{KwInput}{Input}                
			\SetKwInput{KwOutput}{Output} 
			\DontPrintSemicolon
			\textbf{Input:} $\bm{y}$, $\epsilon=10^{-6}$, max iterations ($J$), $\bm{A}_l$, $\bm{A}_s$.\\ 
			\textbf{Initialization:} $\rho=10^{-2}$,~$\rho_{o}=1.001$,~$t=0$,\\$\bm{L}^{0}=\bm{0}_{M,N}$,~$\bm{S}^{0}=\bm{0}_{M,N}$,  $\bm{u}^{0}=\bm{0}_{K}$. 
			\vspace{3pt} 
			
			\While{$\left\Vert \bm{A}_l\text{\normalfont{vec}}(\bm{L})+\bm{A}_s\text{\normalfont{vec}}(\bm{S})-\bm{y} \right\Vert^2_2>\epsilon$~\text{or}~$t<J$}
			{
				\begin{equation*}
					\begin{aligned}
						&\text{Get all the elements of}~ \bm{\lambda}_{LT}^t~\text{by \eqref{eq:lowrank_sparse_6b}, then estimate}\\ &\bm{L}~\text{by,}\\
						&\bm{L}^{t+1} = \text{SVT}_{\bm{\lambda}_{LT}^t}  \left( \bm{A}_l^\ast\left(\bm{y} -\bm{A}_s\text{vec}(\bm{S}^{t})+\dfrac{1}{\rho}  \bm{u}^t\right)\right).\\		
						&\text{Get all the elements of} ~\bm{\lambda}_{ST}^t~\text{by \eqref{eq:lowrank_sparse_6c}, then estimate}\\ &\bm{S}~\text{by,}\\
						&\bm{S}^{t+1} = \text{ST}_{\bm{\lambda}_{ST}^t}\left( \bm{A}_s^	\ast\left(\bm{y} -\bm{A}_l\text{vec}(\bm{L}^{t+1})+\dfrac{1}{\rho}  \bm{u}^t\right)\right).\\
						&\bm{u}^{t+1}=\bm{u}^{t}\!+\!\rho\! \left(\!\bm{A}_s\text{vec}\left(\bm{S}^{t+1}\!\right) \!+\!\bm{A}_l\text{vec}\left(\bm{L}^{t+1})\right)\!-\!\bm{y}\right).\\
						&\rho = \text{min} (\rho_{o} \times \rho,\rho_{\text{m}}),~ \text{and}~t=t+1.\\				
						&\bm{L} \gets \bm{L}^{t+1}~\text{and}~\bm{S} \gets \bm{S}^{t+1}.
					\end{aligned}
				\end{equation*}
			}	
			\KwOutput{$\bm{L}$, $\bm{S}$.}}
		
		\caption{ Low-rank-plus-sparse recovery \\ algorithm}
		\label{alternating_algo_admm}
	\end{algorithm}
	\normalfont
	
	\section{Unfolding ADMM based low-rank plus sparse recovery algorithm}
	\label{sec:Unfolding_ADMM}
\begin{figure}[!t]
		\centering
		\includegraphics[width=1\linewidth]{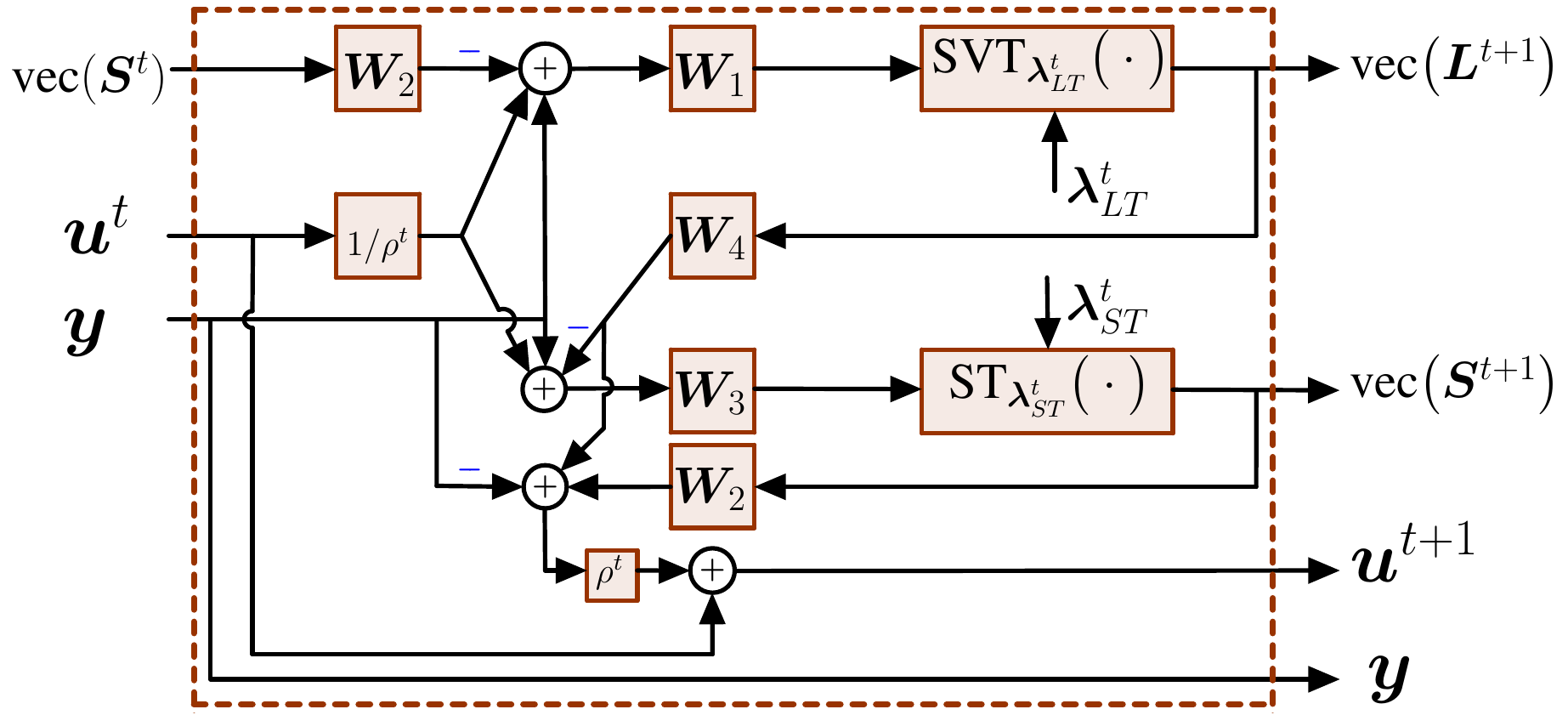}
		\caption{Block diagram of the~\mbox{$t-$th}~layer of the DNN which mimic the low-rank plus sparse recovery Algorithm~\ref{alternating_algo_admm}.}
		\label{fig:dnnls}
	\end{figure}

	In this section, we are going to discuss the ADMM algorithm unfolding using a dense DNN. An iterative algorithm given in Algorithm~\ref{alternating_algo_admm} utilizes a sequence of steps to estimate the solution based on the ADMM steps given in~\eqref{eq:lowrank_sparse_A_WADMM_SVD}, \eqref{eq:lowrank_sparse_A_WADMM_ST} and \eqref{eq:lowrank_sparse_A_ADMM_U}. Here,  previous estimates are used in the next iteration. Thus, this kind of iterative algorithm can be considered as a recurrent neural network. The~\mbox{$t-$th}~iteration of the iterative algorithm is modeled as the \mbox{$t-$th}~layer of the deep neural network. Each matrix multiplication given in the ADMM steps~\eqref{eq:lowrank_sparse_A_WADMM_SVD}, \eqref{eq:lowrank_sparse_A_WADMM_ST} and \eqref{eq:lowrank_sparse_A_ADMM_U} is considered as the weights of the deep neural network. Now, these matrix multiplications are implemented as linear layers i.e., a single matrix multiplication is implemented as a single linear layer without a bias. Here, our main objective is to learn the per iteration weights of the network and thresholding parameters ${\lambda}_{S}$ and ${\lambda}_{L}$ given in~\eqref{eq:lowrank_sparse_6c} and \eqref{eq:lowrank_sparse_6b} from the training data. To this end, the \mbox{$t-$th}~layer of the neural network is represented by the following equations.
	\begin{equation}
		\label{eq:lowrank_sparse_A_WADMM_SVD_NNW}
		\bm{L}^{t+1} = \text{SVT}_{\bm{\lambda}_{LT}^t}\left( \bm{W}_{1}^t\left(\bm{y} -\bm{W}_{2}^t\text{vec}\left(\bm{S}^{t}\right)+\dfrac{\bm{u}^t}{\rho^t}\!\right)\right),
	\end{equation}
	\begin{equation}
		\label{eq:lowrank_sparse_A_WADMM_ST_NNW}
		\bm{S}^{t+1} = \text{ST}_{\bm{\lambda}_{ST}^t}\left(\bm{W}_{3}^t\left(\bm{y} -\bm{W}_{4}^t\text{vec}\left(\bm{L}^{t+1}\right)+\dfrac{\bm{u}^t}{\rho^t}  \right)\right),
	\end{equation}
	\begin{equation}
		\label{eq:lowrank_sparse_A_WADMM_U_NNW}
		\begin{aligned}
			\bm{u}^{t+1} =\bm{u}^{t}+\rho^t \left(\bm{W}_{2}^t\text{vec}\left(\bm{S}^{t+1}\right) +\bm{W}_{4}^t\text{vec}\left(\bm{L}^{t+1}\right)-\bm{y}\right).
		\end{aligned}
	\end{equation}
	Here, $\bm{W}_{1}^t$, $\bm{W}_{2}^t$, $\bm{W}_{3}^t$ and $\bm{W}_{4}^t$ are the weights of the \mbox{$t-$th}~layer of the deep neural network as shown in Fig.~\ref{fig:dnnls}, their initial values are $\bm{W}_{1}^t\!=\!\bm{A}_l^\ast$, $\bm{W}_{2}^t\!=\!\bm{A}_s$, $\bm{W}_{3}^t\!=\!\bm{A}_s^\ast$ and $\bm{W}_{4}^t\!=\!\bm{A}_l$ to mimic the ADMM Algorithm~\ref{alternating_algo_admm}. Further, $\bm{\lambda}_{LT}^t$ and $\bm{\lambda}_{ST}^t$ are the thresholding vectors of the~\mbox{$t-$th}~layer as given in~\eqref{eq:lowrank_sparse_A_WADMM_SVD} and \eqref{eq:lowrank_sparse_A_WADMM_ST}. Note that, $\bm{\lambda}_{LT}^t$ and $\bm{\lambda}_{ST}^t$ depend on the previous estimates of the $\bm{L}^{t}$, $\bm{S}^{t}$ and two parameters ($\lambda_{S}$ and $\lambda_{L}$). Here, we consider matrices $\bm{W}_{1}^t$, $\bm{W}_{2}^t$, $\bm{W}_{3}^t$ and $\bm{W}_{4}^t$ are tied over all the layers, i.e., sharing weights. However, we do not consider thresholding parameters ($\lambda_{S}$ and $\lambda_{L}$) and $\gamma$ to be tied over all layers, i.e., each layer has its own thresholding parameters. To this end, $\Theta=\left\{\lambda_{S}^{t},\lambda_{L}^{t},\gamma^{t}, \rho^t, \bm{W}_{1}, \bm{W}_{2}, \bm{W}_{3}, \bm{W}_{4}\right\}$ represents the set of learning parameters. Here, $\lambda_{S}^{t}$ and $\lambda_{L}^{t}$ are the thresholding parameters of the~\mbox{$t-$th}~layer. The unfolded architecture of the low-rank plus sparse recovery Algorithm~\ref{alternating_algo_admm} is shown in Fig.~\ref{fig:dnnls}. Here, this figure shows a single layer of the deep neural network (\mbox{$t-$th}~layer) and this layer is equivalent to the \mbox{$t-$th}~iteration of  Algorithm~\ref{alternating_algo_admm}.
	\subsection{Training phase}
	\label{sec:Training_ADMM}
	In the training phase, the DNN is trained in a supervised manner. Here, the DNN learns the parameters given in  $\Theta=\left\{\lambda_{S}^{t},\lambda_{L}^{t},\gamma^{t}, \rho^t, \bm{W}_{1}, \bm{W}_{2}, \bm{W}_{3}, \bm{W}_{4}\right\}$. 
	Suppose that the DNN has $T$ layers, then the outputs of the DNN in the training phase for the \mbox{$i-$th}~sample are given by $\hat{\bm{L}}_i$ and $\hat{\bm{S}}_i$, respectively. Note that, in the training phase, the DNN minimizes the normalized mean squared error
	\begin{equation}
		\text{MSE}=\dfrac{1}{T_s}\sum_{i=1}^{T_s}\left(\dfrac{1}{2} \dfrac{\left\Vert\hat{\bm{L}}_i-\bm{L}_i \right\Vert_F^2} { \left\Vert \bm{L}_i \right\Vert_F^2}+\dfrac{1}{2}~\dfrac{\left\Vert \hat{\bm{S}}_i-\bm{S}_i \right\Vert_F^2} { \left\Vert \bm{S}_i \right\Vert_F^2}\right),
		\label{dnnmse}
	\end{equation}
	where $\bm{S}_i$ and $\bm{L}_i$ are \mbox{$i-$th}~ground-truth low rank and sparse matrices and $T_s$ is the number of training samples.
		
	In this work, we consider three versions of the proposed DNN based ADMM  thresholding as follows.
	(a) Parameter learning with non-adaptive thresholding (i.e., $g_s(x)=g_l(x)=1$). This approach is named as ADMM based trained RPCA with thresholding \mbox{(TRPCA-T)}. For the parameter learning with adaptive thresholding, we consider two versions based on two decay functions as described in \ref{EWST}. These two approaches are named as follows. (b) ADMM based trained RPCA with adaptive thresholding based on logarithm heuristic \mbox{(TRPCA-AT(log))}. (c) ADMM based trained RPCA with adaptive thresholding based on exponential heuristic \mbox{(TRPCA-AT(exp))}.
	
	To have a comparison with our proposed approach, we consider two approaches. In the first approach, we consider the untrained ADMM approach to solve the low-rank plus sparse recovery as given in Algorithm~\ref{alternating_algo_admm} with $g_s(x)\!=\!g_l(x)\!=\!1$ (i.e., by using the $\ell_1-$norm and nuclear norm). This method is named as ADMM based untrained RPCA with thresholding \mbox{(URPCA-T)}. As a second option, we consider the convex relaxation of the low-rank plus sparse recovery problem given in~\eqref{eq:lowrank_sparse_A_cov}. This method is named as low-rank plus sparse recovery with convex relaxation \mbox{(LRPSRC)}.
	\subsection{Computation complexity}
	In this sub-section, the computational complexity of the proposed DNN is discussed. Note that the $t-$th iteration of Algorithm~\ref{alternating_algo_admm} is shown in Fig. \ref{fig:dnnls}. Here, a single layer of the DNN consists of four dense linear layers and their weight matrices are given with $\bm{W}_{1}^t,~\bm{W}_{3}^t \in~\mathbb{C}^{K\times MN}$ and $\bm{W}_{2}^t,~\bm{W}_{4}^t\in~\mathbb{C}^{MN\times K}$.
		In the feed-forward propagation, data propagation is given in eqs.~\eqref{eq:lowrank_sparse_A_WADMM_SVD_NNW}, \eqref{eq:lowrank_sparse_A_WADMM_ST_NNW} and \eqref{eq:lowrank_sparse_A_WADMM_U_NNW}. Now, for $T_s$ number of training samples and $n$ number of epochs, the computational complexity of the feed-forward propagation is  $\mathcal{O}(6T_sn(MNK)+ \mathcal{O}(T_sn(M^2N+MN^2+N^3))+ \mathcal{O}(T_sn(M^2N+N^2M))~\approx\mathcal{O}(T_sn(MNK+M^2N+N^2M+N^3))$. When $M=N$, the computational complexity of the feed-forward propagation is given by $\mathcal{O}(T_sn(N^2K+N^3))$. Here, $\mathcal{O}(\cdot)$ is the Big O notation for asymptotic computational complexity analysis \cite{chivers2015introduction}. It is used to analyze the computational performance of an algorithm as input size  increases. For the back propagation, the computational complexity of the linear layers is given by $\mathcal{O}(6T_sn(MNK)$ and for the back propagation through SVD it is given by $\mathcal{O}(T_sn(M^2N+MN^2+N^3))+ \mathcal{O}(T_sn(M^2N+N^2M))$. Hence, the training complexity of the DNN is the addition of the feed-forward propagation and the back propagation complexities. Now, for $M=N$, the training complexity of the DNN is given by $\mathcal{O}(2T_sn(N^2K+N^3))~\approx\mathcal{O}(T_sn(N^2K+N^3))$. This computational complexity corresponds to the single iteration of Algorithm \ref{alternating_algo_admm}. Now, for $T$ iterations/layers the training computational complexity is given by $\mathcal{O}(TT_sn(N^2K+N^3))$. The testing computational complexity is the feed-forward propagation complexity of the data through the DNN. It is given by $\mathcal{O}(TN_s(N^2K+N^3))$, here $N_s$ is the number of testing samples.
	\section{Results}
	\label{sec:results}
	In this section numerical results are presented. First, the performance of deep learning-based trained ADMM adaptive thresholding is evaluated with a generic real-valued Gaussian model and next a complex-valued SFCW radar model given in Section \ref{sec:SFCWSystem_Model} is used.	
	\subsection{Generic Gaussian model}
	\label{sec:resultsgaussian}
	Here, the elements of the matrix~$\bm{A}_l\!=\!\bm{A}_s\!=\!\bm{A}~\in~\mathbb{R}^{K\times MN}$ are generated once from an i.i.d. Gaussian distribution with zero mean and unit variance. $\bm{y}$ is generated based on the~\eqref{eq:SM_G}. Further, we normalized the matrices $\bm{S}$ and $\bm{L}$ to have a unit Frobenius norm (i.e., $\left\Vert \bm{L} \right\Vert_F^2=\left\Vert \bm{S} \right\Vert_F^2=1$). The signal-to-noise ratio is $\text{SNR}:=\left\Vert\bm{A}\text{vec}(\bm{L}+\bm{S})\right\Vert_2^2\big{\slash}\left\Vert\bm{n}\right\Vert_2^2=20$dB and $M=30$, $N=30$.  Generally, many training samples are required to train a deep neural network. However, due to the specific architecture of the iterative algorithm, here we are able to train the DNN with a small data set of $500$ samples. In the training phase, the adaptive moment estimation (Adam) optimizer \cite{kingma2017adam} with a learning rate of $0.1$ is used to train the DNN and batch size was set as $50$. Note that, in the Gaussian model, the DNN only learns the $\lambda_{S}^{t}$ and $\lambda_{L}^{t}$ instead of all the parameters given in $\Theta$. This is due to the fact that the performance gain improvement learning all the parameters given in $\Theta$ is very small compared to learning only $\lambda_{S}^{t}$ and $\lambda_{L}^{t}$. Here, we initialize $\bm{W}_{1}^t=\bm{A}_l^\dagger$, $\bm{W}_{2}^t=\bm{A}_s$, $\bm{W}_{3}^t=\bm{A}_s^\dagger$ and $\bm{W}_{4}^t=\bm{A}_l$ to mimic the ADMM Algorithm~\ref{alternating_algo_admm} and $\gamma=1$. The DNN is trained over a maximum of $500$ epochs. In the inference phase, to evaluate the performance of the DNN, the normalized average root mean squared error is used. For low-rank and sparse matrix, it is given by
	\begin{align}
		\text{RMSE}_\text{L}&=\dfrac{1}{N_s}\sum_{i=1}^{N_s}\left( \dfrac{\left\Vert\hat{\bm{L}}_i-\bm{L}_i \right\Vert_F }{\left\Vert \bm{L}_i \right\Vert_F}\right),\\
		\text{RMSE}_\text{S}&=\dfrac{1}{N_s}\sum_{i=1}^{N_s}\left( \dfrac{\left\Vert \hat{\bm{S}}_i-\bm{S}_i \right\Vert_F}{\left\Vert \bm{S}_i \right\Vert_F} \right).
		\label{dnnmse}
	\end{align}
	The CRB given in \eqref{eq:crb} is based on the combined recovery error of both low-rank and sparse matrices. Now, for that, first we vectorize the matrices $\bm{L}$ and $\bm{S}$ and next, we cascade them to generate a single vector. For the $i-$th ground truth and estimated low-rank and sparse matrices these vectors are given by $\bm{x_i}=[\text{vec}(\bm{L}_i)^T~ \text{vec}(\bm{S}_i)^T]^T$ and $\bm{\hat{x}_i}=[\text{vec}(\hat{\bm{L}}_i)^T~ \text{vec}(\hat{\bm{S}}_i)^T]^T$, respectively. Now, the combined average root mean squared error is given by
	\begin{align}
		\text{RMSE}_\text{LS}&=\dfrac{1}{N_s}\sum_{i=1}^{N_s}\left( \left\Vert\bm{x_i} -\bm{\hat{x}_i} \right\Vert_2\right).
		\label{dnnmsecrb}
	\end{align}
	\begin{figure}[!t]
		\centering
		\includegraphics[width=0.95\linewidth]{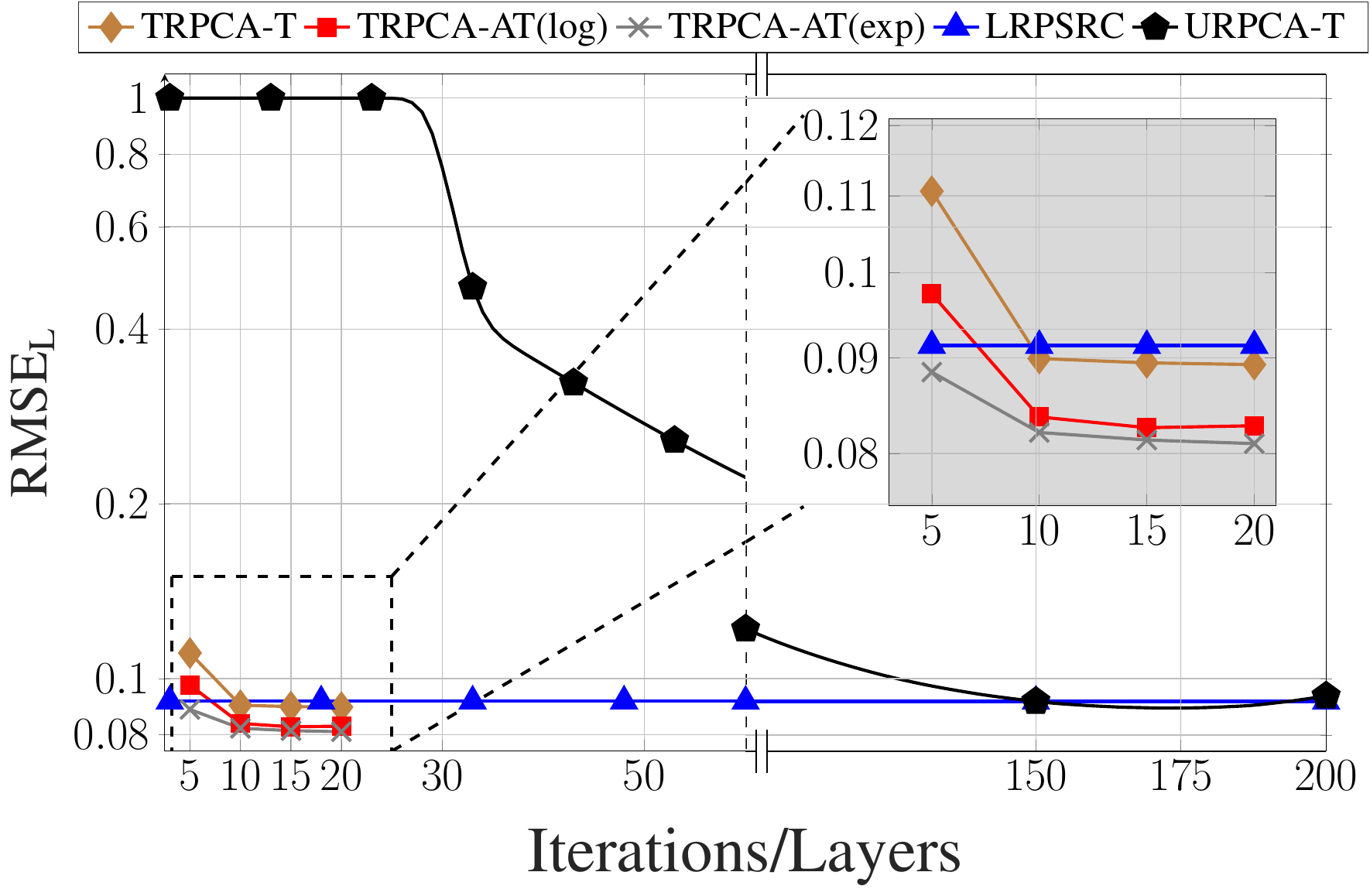}
		\includegraphics[width=0.95\linewidth]{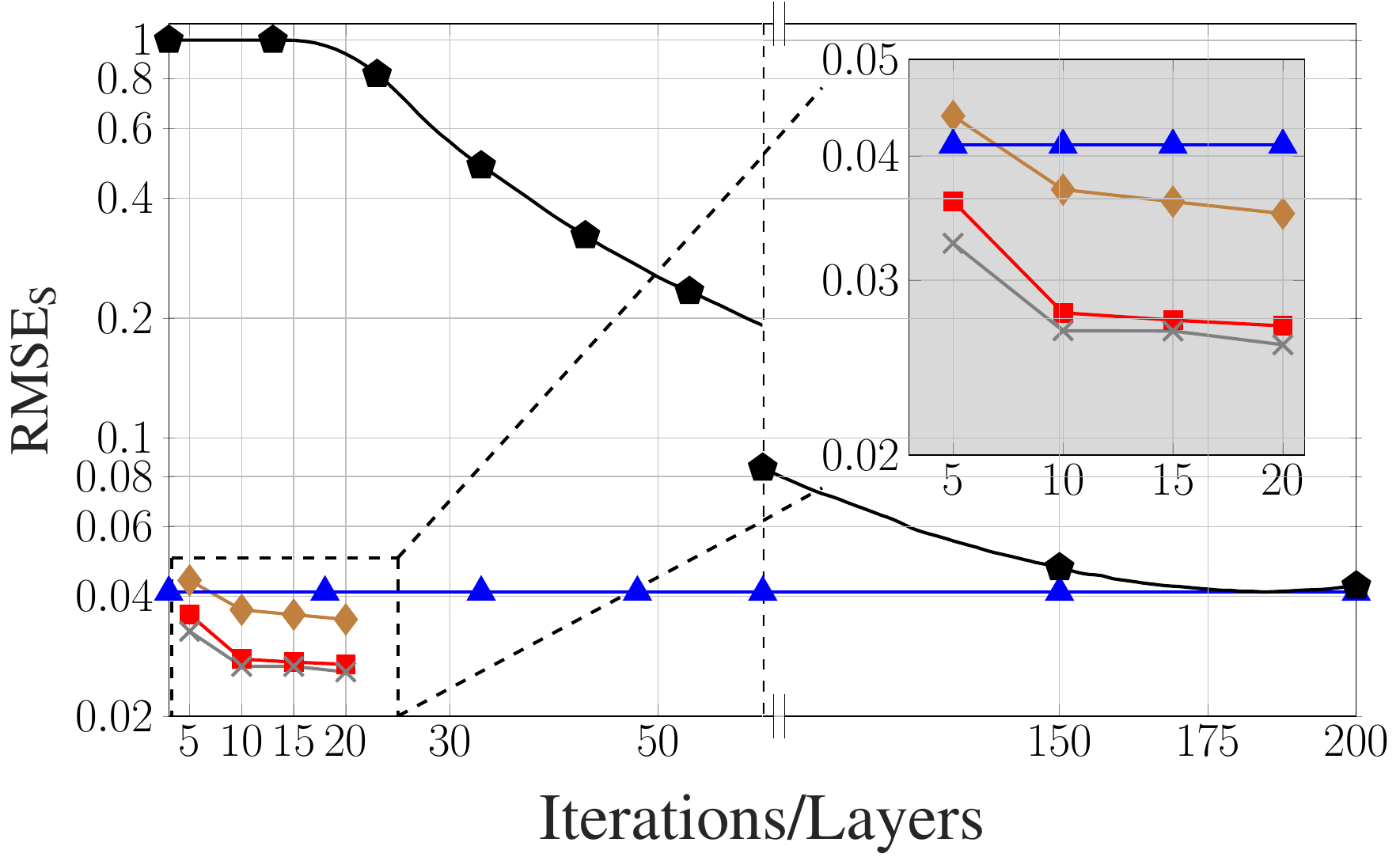}
		\caption{Average  recovery error of low-rank (top) and sparsity (bottom) contributions for compression ratio $K{\slash}MN=50\%$ .}
		\label{fig:lsrecovery50}
	\end{figure}
	\begin{figure}[!t]
		\centering
		\includegraphics[width=0.85\linewidth]{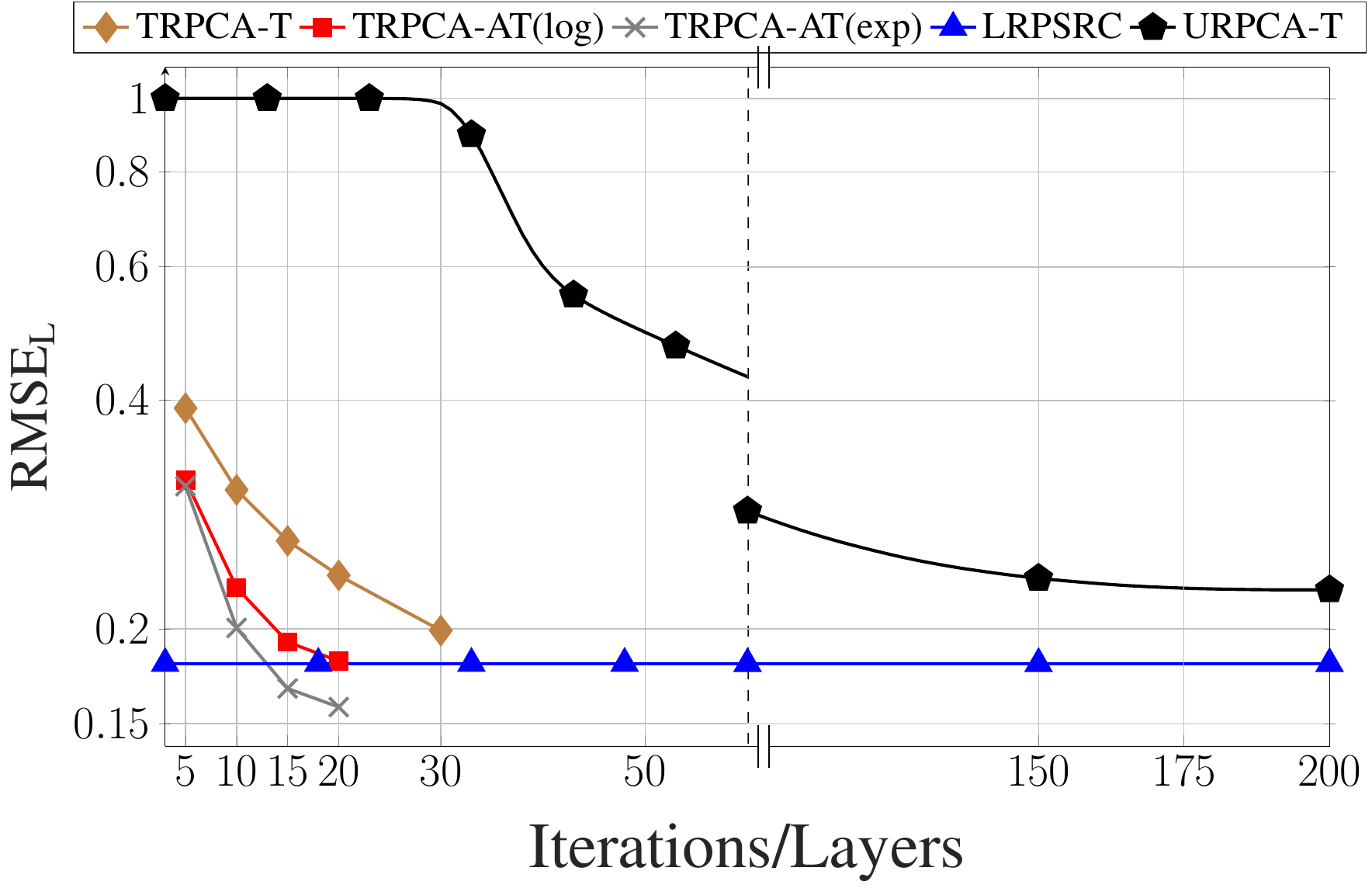}
		\includegraphics[width=0.85\linewidth]{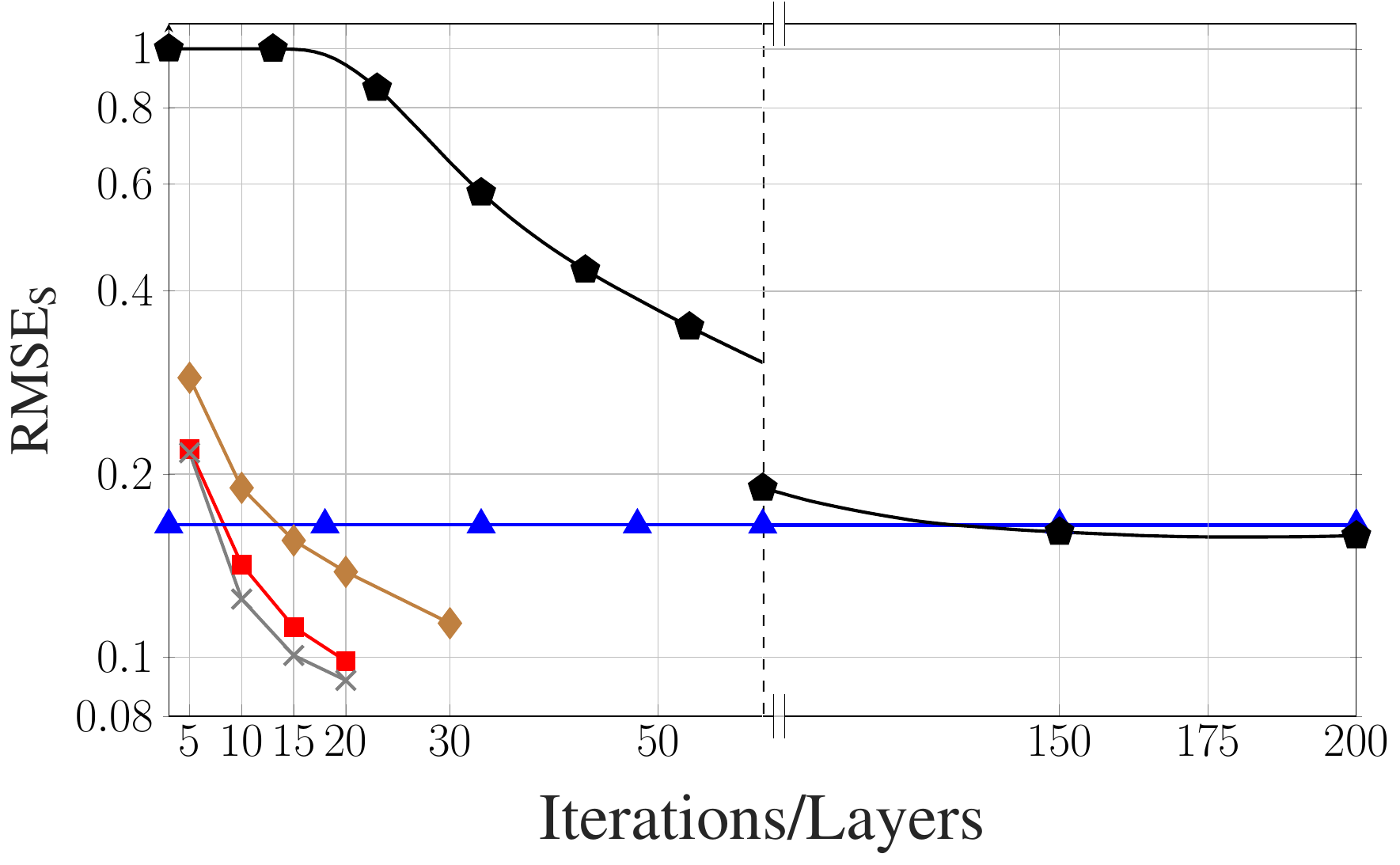}
		\caption{Average  recovery error of low-rank (top) and sparsity (bottom) contributions for compression ratio $K{\slash}MN=25\%$ .}
		\label{fig:lsrecovery25}
	\end{figure}
	\begin{figure}[!t]
		\centering
		\includegraphics[width=0.85\linewidth]{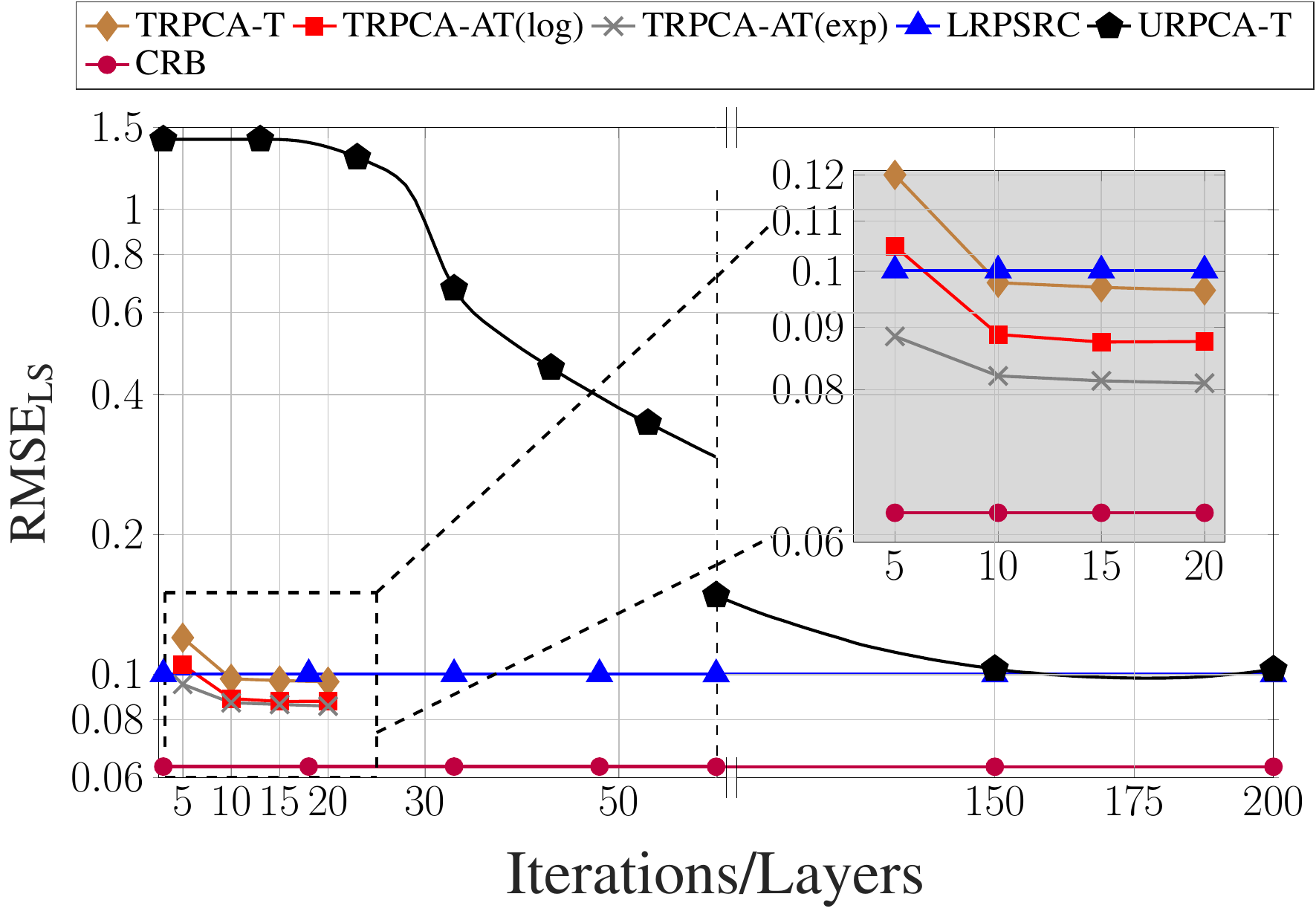}
		\includegraphics[width=0.85\linewidth]{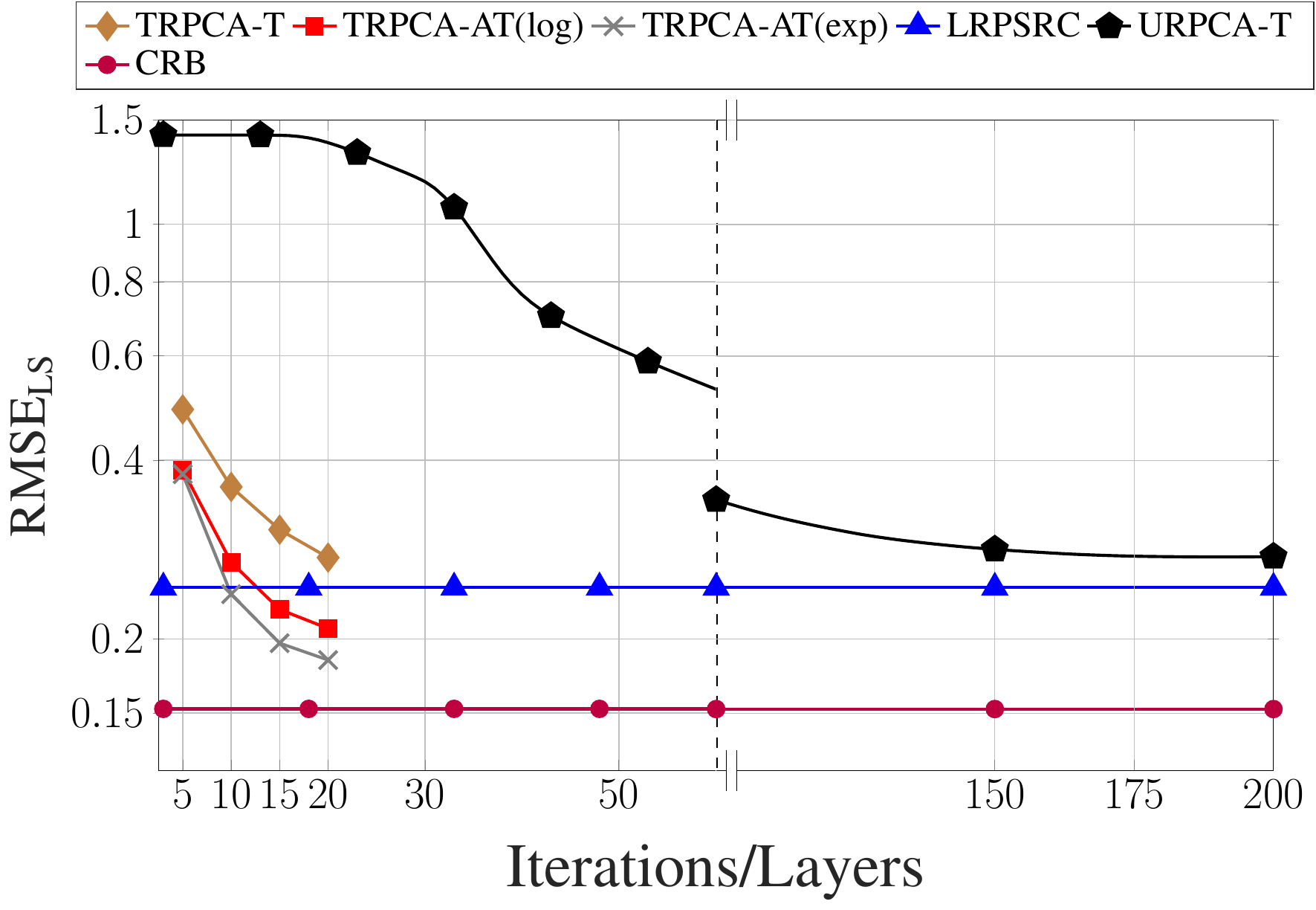}
		\caption{Average combined recovery error of low-rank and sparse matrices as given in~\eqref{dnnmsecrb} for compression ratio $K{\slash}MN=50\%$ (top) and $K{\slash}MN=25\%$ (bottom).}
		\label{fig:lsrecoveryCRB}
	\end{figure}\\
	\noindent The outputs of a DNN with $T$ layers is for the \mbox{$i-$th}~testing sample are given by $\hat{\bm{S}}_i$ and $\hat{\bm{L}}_i$, respectively. Here, $\bm{S}_i$ and $\bm{L}_i$ are \mbox{$i-$th}~ground-truth low rank and sparse matrices and $N_s=1200$ is the number of testing samples. In this work, training and testing data are synthetically generated based on the system model given in~\eqref{eq:SM_G}. Therefore, ground-truth low-rank and sparse matrices are available. In case only the received data vector~$\bm{y}$ in~\eqref{eq:SM_G} is available, Algorithm~\ref{alternating_algo_admm} can be used to generate ground-truth low-rank and sparse matrices.
	Both Algorithm~\ref{alternating_algo_admm} and LRPSRC given in~\eqref{eq:lowrank_sparse_A_cov} are implemented using Matlab \cite{MATLAB:2019} and LRPSRC is solved using the CVX package \cite{cvx}. Notice that, in the LRPSRC $\lambda_l$ and $\lambda_s$ are set to $1$, and $1\big{\slash}\sqrt{\text{max}(M,N)}$, respectively as suggested by \cite{candes2011robust}. Note that for Algorithm~\ref{alternating_algo_admm}, there is no specific rule to select the $\lambda_{l}$ and $\lambda_{s}$ and $\rho$, thus they are manually tuned based on the data\footnote{When $\bm{A}$ is identity matrix, there is a specific rule to select $\lambda_{s}$ as $1\big{\slash}\sqrt{\text{max}(M,N)}$ \cite{candes2011robust}.}. Note that as a rule of thumb,
	thresholding parameters ${\lambda}_{S}$ and ${\lambda}_{L}$ given in~\eqref{eq:lowrank_sparse_6c} and \eqref{eq:lowrank_sparse_6b} are initialized as ${\lambda}_{S}={\lambda}_{s}/\rho$ and ${\lambda}_{L}={\lambda}_{l}/\rho$, respectively \cite{candes2011robust}. The Pytorch package is used to implement the DNN \cite{NEURIPS2019_9015} .
	
	 The average normalized RMSEs for the different number of layers of the DNN are for compression ratio ($K{\slash}MN$) $50\%$ and $25\%$ shown in Figs.~\ref{fig:lsrecovery50}~and~\ref{fig:lsrecovery25}, respectively. Figs.~\ref{fig:lsrecovery50}~and~\ref{fig:lsrecovery25} show that the proposed DNN based thresholding (\mbox{TRPCA-AT(log)} and \mbox{TRPCA-AT(exp)}) outperforms the \mbox{URPCA-T} and the \mbox{LRPSRC}. Further, it is observed that as the number of layers increases the average RMSE decreases. For the $50\%$ compressing ratio, the average RMSE does not show a large variance after ten layers. However, for the $25\%$ compressing ratio this is not the case. This is due to the fact that, as the compression ratio increases recovering of the low-rank and the sparse matrices become more challenging.
	 
	  Further, the \mbox{TRPCA-AT} outperforms the \mbox{TRPCA-T}. This performance improvement is mainly due to the iterative reweighting of  $\ell_1-$norm and nuclear norm minimization. Also, the improvement over non-weighting to iterative~rewei\-ghting is more visible as the compression ratio increases (i.e., as the problem getting more challenging). As an example, for the $25\%$ compressing ratio, the average \mbox{RMSE} improvement between the \mbox{TRPCA-T} with twenty layers and \mbox{TRPCA-AT(exp)} with twenty layers for the low-rank and sparse components are $32.93\%$ and $50.77\%$, respectively. However, for the $50\%$ compressing ratio, this improvement for the low-rank and sparse components are $9.31\%$ and $26.21\%$, respectively. Further, we observe slight performance gains as the decay function is changed from \mbox{log-determinant} to exponential. As we compare the number of layers of the \mbox{TRPCA-AT} to the number of iterations of the \mbox{URPCA-T}, the \mbox{TRPCA-AT} achieves $1:20$ improvement. In more details, \mbox{TRPCA-AT} with ten layers outperforms \mbox{URPCA-T} with $200$ iterations. Therefore, in the testing phase (inference phase), our proposed approach (\mbox{TRPCA-AT}) is twenty times faster than the conventional untrained approach (\mbox{URPCA-T}).
	  
	Fig. \ref{fig:lsrecoveryCRB} shows the CRB of the
		combined low-rank and sparse matrices estimation for compression ratios $50\%$ and $25\%$. It can be
		seen that the non-convex approach \mbox{TRPCA-AT} has the closest performance to the CRB. As the compression ratio increases, the estimation of low-rank and sparse matrices from compressive measurements becomes more challenging. This can be seen as the compression ratio changes from $50\%$ to $25\%$ the CRB has increased.
	\subsection{SFCW radar model}
	In this subsection, the performance evaluation of the ADMM based trained RPCA with adaptive thresholding was performed based on the SFCW radar model given in Section \ref{sec:SFCWSystem_Model}. In the simulations, we set the carrier frequency $f_c$ of $300$~GHz and 
	bandwidth $B$ as $5$~GHz. Further, both $N$ and $M$ are set as $30$. The inter-antenna spacing  is chosen as half of the wavelength of $f_c$. Here, we consider a single-layered structure and the distance to the front surface of the layered structure is $1.0$~m. Both height and length of the layered structure is $0.5$~m. In the simulations, we consider six defects and the scene is partitioned into a $16 \times 16$ grid with equal grid size (i.e., $Q=256$). The grid size is selected according to the Rayleigh resolution of the radar. 
	In the simulations, the signal-to-noise ratio is $\text{SNR}\!:=\!\left\Vert\bm{\Phi}\text{vec}(\bm{Y}^l)\!+\!\bm{\Phi}\bm{D}\bm{s}\right\Vert_2^2\big{\slash}\left\Vert\bm{\Phi}\text{vec}(\bm{Z})\right\Vert_2^2\!=\!20$dB. Note that the SFCW data consists of complex numbers, thus, in this work, we implemented the DNN which supports complex numbers by using the PyTorch version $1.8.1$. Here, we initialize $\bm{W}_{1}^t=\bm{A}_l^H$, $\bm{W}_{2}^t=\bm{A}_s$, $\bm{W}_{3}^t=\bm{A}_s^H$ and $\bm{W}_{4}^t=\bm{A}_l$ to mimic the ADMM Algorithm~\ref{alternating_algo_admm}.
	
	Interestingly, in contrast to the generic Gaussian model only learning the $\lambda_{S}^{t}$ and $\lambda_{L}^{t}$ does not achieve satisfactory average RMSEs of the low-rank and sparse components. Therefore, we enable learning all parameters given in $\Theta$. Further, we notice that the stochastic gradient descent
	(SGD) optimizer \cite{sutskever2013importance} performs better than the Adam optimizer in learning all the parameters given in $\Theta$ together. Therefore, we consider a three-stage training process for better learning.
	\begin{figure}[!t]
		\centering
		\includegraphics[width=0.85\linewidth]{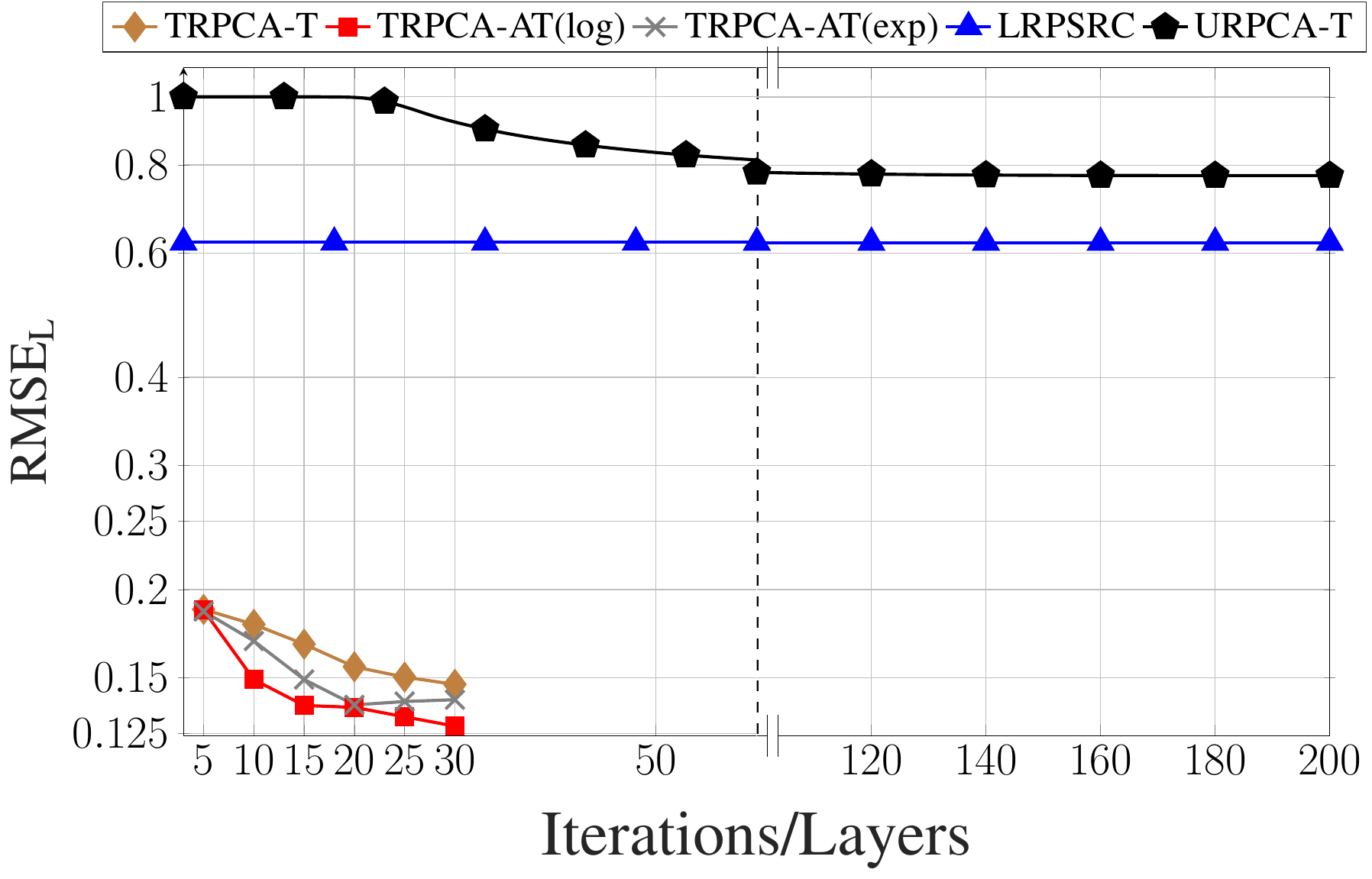}
		\includegraphics[width=0.85\linewidth]{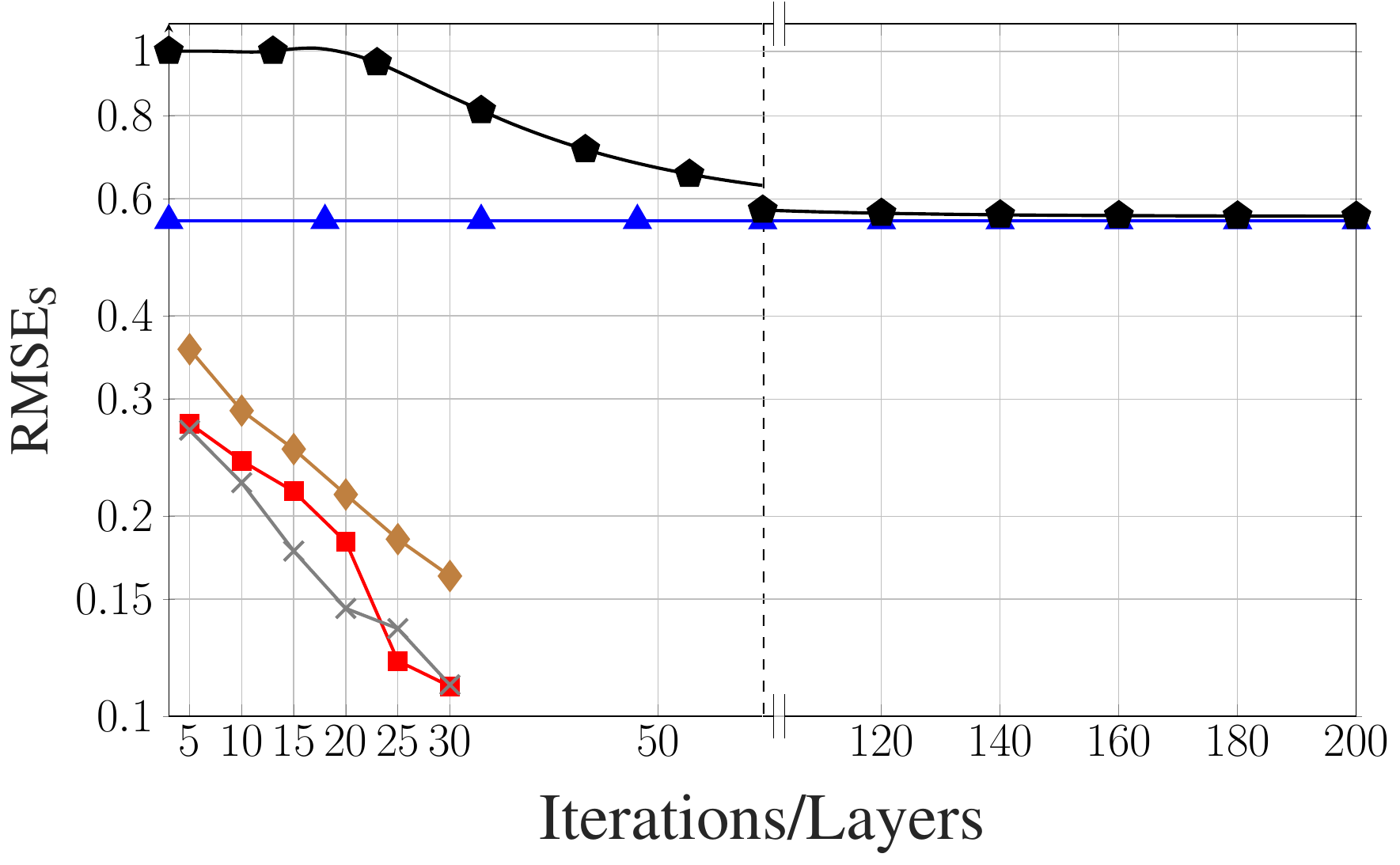}
		\caption{Average RMSE of low-rank (top) and sparsity (bottom) contributions for $K{\slash}MN=20\%$ for SFCW model.}
		\label{fig:lsrecovery_sar}
	\end{figure}
	\begin{figure*}[!t]
		\centering
		\includegraphics[width=1\linewidth]{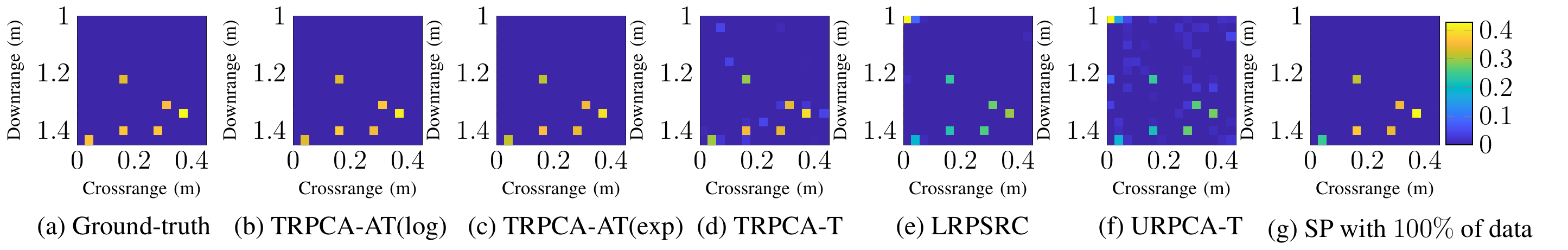}
		\caption{Object recovery in single case for compression ratio $K/MN=20\%$ }
		\label{fig:sar20}
	\end{figure*}
	
	In the first stage, we only learn the $\lambda_{S}^{t}$ and $\lambda_{L}^{t}$ for $50$ epochs using the Adam optimizer. In the second stage, we learn all parameters given in $\Theta$ using the SGD optimizer for $50$ epochs. Finally, we only learn the $\lambda_{S}^{t}$ and $\lambda_{L}^{t}$ for $15$ epochs using the Adam optimizer. Also, we slightly adjusted the learning rate as the number of layers of the DNN increases. For the first stage, we employed learning rates $1\times10^{-1}$, $5\times10^{-2}$ and  $5\times10^{-3}$ for the  DNN with $5$\slash$10$, $15$\slash$20$ and $25$\slash$30$ layers, respectively. Here, only exception is the TRPCA-T. For the TRPCA-T, we consider the learning rate of $5\times10^{-2}$ for the DNN with $25$\slash$30$ layers as well. This is due to the fact that, the non-adaptive thresholding based TRPCA-T is less sensitive to the change of parameters compared to the  adaptive thresholding based TRPCA-AT. For the second and third stages, we employed the learning rates $1\times10^{-3}$, $2.5\times10^{-4}$ for all layers combinations of the DNN, respectively. The main reason to consider the third training stage is that, it achieves higher performance gains with respect to continuation of the second training stage for another $15$ epochs. Also, note that there is a specific reason to use the first stage without directly using the second stage. This is due to the unbalance of $\bm{A}_s$ and $\bm{A}_l$ of the SFCW model compared to the generic Gaussian model. Note that, $\bm{A}_s=\bm{\Phi}\bm{D}$ and $\bm{A}_l=\bm{\Phi}$. More specifically, $\bm{A}_l=\bm{\Phi}$ and $\bm{\Phi}$ is the selection matrix. This matrix has a single \mbox{non-zero} element of value $1$ in each row to indicate the selected frequency of a particular antenna if that antenna is selected. However,~$\bm{A}_s=\bm{\Phi}\bm{D}$ so,~$\bm{A}_s$ is a combination of the selection matrix and the matrix~$\bm{D}$. Note that, the matrix~$\bm{D}$ is generated based on the time delays of the grid (as described in Section \ref{sec:SFCWSystem_Model}). Therefore, the matrix~$\bm{A}_s$ contains more information compared to the matrix~$\bm{A}_l$. Therefore,~$\bm{A}_s$ is more difficult to learn compared to~$\bm{A}_l$. This results in an imbalance in the training phase if we directly start with the stage two. That means the RMSE of the low-rank component tends to be much smaller compared to the RMSE of the sparse component in the training phase if we directly start with the second stage.
	
	 In the defect detection by SFCW radar, we consider a data set of $600$ samples. Here, $500$ data samples are used for training and validation and $100$ data samples are used for testing. Here, we used Matlab \cite{MATLAB:2019} to generate the SFCW data based on the system model given in~\eqref{eq:SS_eq_rx_all_cs}. The average normalized RMSEs for the different numbers of layers of the DNN for the compression ratio ($K{\slash}MN$) $20\%$ is shown in Fig.~\ref{fig:lsrecovery_sar}. The figure shows that the proposed \mbox{TRPCA-AT} outperforms both \mbox{URPCA-T} and the \mbox{LRPSRC} given in~\eqref{eq:lowrank_sparse_A_cov}. Further, in terms of the average RMSE, the \mbox{TRPCA-AT} and \mbox{TRPCA-T} with five layers outperform \mbox{URPCA-T} with $200$ iterations. Therefore, as we compare the number of layers of the \mbox{TRPCA-AT} to the number of iterations of the \mbox{URPCA-T}, the \mbox{TRPCA-AT} achieves an $1:40$ improvement for the SFCW radar data, i.e., our proposed approach (\mbox{TRPCA-AT}) is forty times faster than the conventional untrained approach (\mbox{URPCA-T}). Moreover, based on the results shown in Fig.~\ref{fig:lsrecovery_sar}, the \mbox{TRPCA-AT} shows better performance compared to the \mbox{TRPCA-T}. Also, note that with $20\%$ compression ratio, the estimation of~$\bm{Y}^l$~and~$\bm{s}$ from~$\bm{y}_{cs}$  in~\eqref{eq:SS_eq_rx_all_cs} is more challenging. Therefore, the average RMSE of the \mbox{LRPSRC} is higher than $0.5$. However, the DNN based \mbox{TRPCA-AT} is able to achieve average RMSE in the range of $0.1$ for both sparse and the low-rank components. Since the matrices $\bm{A}_s$ and $\bm{A}_l$ are unequal in the SFCW radar model, we did not consider the CRB benchmark given in \eqref{eq:crb}.
	 
	Next, to further illustrate the defect detection, an image of the recovered detects are formed. Here, the recovered vector~$\bm{s}$ is reshaped into a matrix to obtain an image of the detects as shown in Fig.~\ref{fig:sar20} for a single data sample. As a benchmark here we consider the state-of-the-art subspace projection (SP) \cite{khan2010background} method with the full data set (i.e., compression ratio ($K{\slash}MN)=100\%$). Further, for SP, it is assumed that the number of defects is known. Fig.~\ref{fig:sar20}~(a) shows the actual defect locations. The recovered locations of the defects for 
	the ADMM based trained RPCA with adaptive thresholding based on logarithm heuristic (\mbox{TRPCA-AT(log)}), ADMM based trained RPCA with adaptive thresholding based on exponential heuristic (\mbox{TRPCA-AT(exp)}), ADMM based trained RPCA with thresholding (\mbox{TRPCA-T}), the \mbox{LRPSRC}, ADMM based untrained RPCA with thresholding (\mbox{URPCA-T}) and the SP are shown in Fig.~\ref{fig:sar20}~(b), (c), (d), (e), (f) and (g), respectively. It can be seen that the proposed \mbox{TRPCA-AT} approaches are able to identify all six defects. Further, the proposed \mbox{TRPCA-AT} approaches are able to estimate amplitudes of the recovered defects (vector~$\bm{s}$) closer to the actual defects. Therefore, it is observed that defect detection with the proposed \mbox{TRPCA-AT} approaches outperform state-of-the-art SP even with $20\%$ of data.
	
	\section{Conclusion}
	\label{sec:concl}
	This paper presents a deep learning-based low-rank plus sparse recovery approach for the detection of material defects based on compressive sensing. To this end, an iterative algorithm is developed based on ADMM to estimate the low-rank and sparse contributions with the iterative reweighted nuclear and $\ell_1-$norm minimization. In particular, we propose deep learning based algorithm unfolding to improve the accuracies of the recovered low-rank and sparse components and the speed of convergence of the algorithm.
	The results show that, the deep learning based algorithm unfolding performs substantially better compared to the untrained iterative algorithm in terms of low-rank and sparse component recovery and convergence speed. In addition to that, it is observed that as the compression ratio increases, the improvement with deep learning becomes more visible.

	\bibliographystyle{IEEEtran}
	\bibliography{ref_m}

\end{document}